\documentstyle[preprint,aps]{revtex}
\def \bfgr #1{ \mbox {{\boldmath $#1$}}}
\def \dfrac #1#2 {\displaystyle\frac{#1}{#2}}

\begin{document}
\draft
\def\theequation{\thesection.\arabic{equation}}

\title{
\rightline{DFUPG 92/94}
\vspace*{.5cm}
The Spin-Dependent Structure
Functions of Nuclei in the  Meson-Nucleon Theory}
\author{L.P. Kaptari
\footnote{On leave from
 Bogoliubov's Laboratory of  Physics, JINR, Dubna 141980, Russia.}
, A.Yu. Umnikov \footnote{ On leave from   Theoretical Physics Institute,
University of Alberta, Edmonton,
Alberta
\phantom{$^\dagger\; $}T6G 2J1, and TRIUMF, 4004 Wesbrook Mall,
Vancouver,
B.C. V6T 2A3, Canada}
C. Ciofi degli Atti, and S. Scopetta}
\address{Department of Physics, University of Perugia, and
Istituto Nazionale di
Fisica Nucleare, Sezione di Perugia, Via A.Pascoli, I-060100
Perugia, Italy }
\author{K.Yu. Kazakov\footnote
{On leave from Far-Eastern State University, Vladivostok, Russia.}}
\address{ Theoretical Physics Institute, University of Alberta, Edmonton,
Alberta T6G 2J1, and TRIUMF, 4004 Wesbrook Mall, Vancouver,
B.C. V6T 2A3, Canada.}
\maketitle

\vskip 2cm
\begin{abstract}
A theoretical  approach to the investigation of spin-dependent
structure functions in deep inelastic scattering of polarized leptons
off polarized nuclei,
based
on the effective meson-nucleon theory and
operator product expansion method,
is proposed and applied to deuteron and $^3He$.
The explicit forms of the moments of the deuteron and $^3He$
spin-dependent
structure functions are found  and numerical estimates of the
influence
of nuclear structure effects are presented.
\end{abstract}


\narrowtext
\section{Introduction}
 The large body of experiments
  on deep inelastic lepton--nucleon
 scattering performed in the 70-th and 80-th,
 which resulted in the amazing agreement of the data with the
 quark-parton model and the confirmation of the phenomenon of
 logarithmic violation of
 Bjorken scaling, provided the basis for the
 foundation of Quantum Chromodynamics (QCD) as the theory of
 strong interactions.
 However, the latest experimental results in this field
seem to display deviations from the
 predictions of the quark-parton model (``free QCD'')
 and perturbative QCD.
(\cite{spin}, \cite{nmcgot}).
In this context,
it should be stressed that basic theorems (sum rules) of QCD
require the knowledge of
the spin structure function (SSF) $g_1^n(x)$ of the neutron.

This has been recently extracted by
deep inelastic scattering of polarized leptons off polarized nuclear
targets, e.g. deuteron (Spin Muon Collaboration (SMC)~\cite{smc})
and $^3He$ (E142 experiment ~\cite{slac}) and
these data, combined with earlier data of the
European Muon Collaboration (EMC) on the proton~\cite{spin},
are being currently used to check the predictions of QCD
{}~\cite{CloseR,EllisK}.

The Bjorken sum rule has been computed from E142 data and,
using the result
$\int_0^1g_1^p(x)dx = 0.126 \pm 0.010 \pm 0.015$
obtained by the EMC collaboration ~\cite{spin},
the
value $\int\limits_0^1\, \left (g_1^p-g_1^n \right )\,dx =0.148\pm
0.022.$~\cite{slac} has been found,
to be compared with the theoretical prediction of
$0.187 \pm 0.004$.
At the same time, the
SMC Collaboration estimated for the first time
the first moment of SSF of the ``isoscalar'' nucleon :
${\sf M}_n \; \equiv \int\limits_0^1 g_1^N\,dx =0.023\pm 0.02\pm
0.015,$
and combining again these data with the EMC proton data,
for the Bjorken sum rule was found to be
$\int\limits_0^1\, \left (g_1^p-g_1^n \right )\,dx =0.20\pm
0.05\pm 0.04.$

 New experiments are planned at DESY\cite{desy},
 CERN\cite{smc} and SLAC \cite{slac}, also aimed at an improved
measurement of this fundamental prediction of QCD.

    It is
   worth stressing here that,
  all information on the neutron SSF
   have been and will be obtained
   by
analyzing DIS off polarized
{\em nuclear} targets,
   in particular $^2H$ and $^3He$.
   Therefore such an information
can in principle depend upon
nuclear structure effects, which however are considered
to provide only a minor correction.

As a matter of facts,
the possibility to measure the near free isoscalar nucleon structure
functions by using polarized $^2H$ target is motivated by
the fact that typical nuclear effects in the deuteron
are small and are predominantly determined by
well-known spin-orbital structure of the
deuteron wave function (c.f. Ref.~\cite{fs1,kubj}).
At the same time, the interest in polarized $^3He$ targets
stems
from the observation ~\cite{fri} that since the dominant component
of the $^3He$ wave function is fully symmetric $S$ wave with
the two protons in a spin singlet state, a polarized $^3He$
can be associated to a large extent with a polarized neutron.
It should
however
be pointed out that the precision required by a significant check
of the Bjorken Sum Rule would demand
a quantitative estimate of all possible nuclear effects.
For example, in case of $^3He$, the
$S'$ and $D$--wave components of the small realistic three
body wave function
generate a proton contribution to the
$^3He$ polarization and asymmetry,
which has to be subtracted in order to obtain information on the neutron
properties. Recently, the proton contribution to the process
$\vec{^3He}(\vec e, e')X$ has been quantitatively evaluated by
full convolution approach, where not only Fermi motion
{}~\cite{vol} but binding as well~\cite{ciofi2} were taken into account by
generalizing
the usual convolution approach
(see e.g. Ref.
{}~\cite{ciofi3}), by introducing the concept of spin dependent spectral
function ~\cite{ciofi1,sauer}, and nuclear
effects have been indeed found to be small.
Relativistic light cone calculations have been performed in
Ref. \cite{coe} using the spin dependent spectral function of Ref.
\cite{sauer}; the obtained results practically do not differ
from the non relativistic ones obtained in Ref. \cite{ciofi1}.

The underlying reason for the applicability of the convolution
model is the existence of two typical momentum scales in deep
inelastic processes, which leads to the factorization of the
amplitude of the reaction into two pieces, depending
respectively on the
``large" external
momentum, i.e.
the structure function of the nucleon, and the ``low"
typical momenta
of nucleons in the nucleus, i.e. the momentum distribution of the
nucleons~\cite{jaffeb}. A rigorous method to analyze such a
factorization is based upon the Wilson operator product expansion
(OPE).
A theoretical approach to investigate
deep inelastic scattering on nuclei
by using the OPE
method within the effective meson-nucleon theory
with one--boson--exchange (OBE) interaction has been
suggested in Ref.~\cite{kaz,ukhanna}; these
calculations
have been performed within well--defined approximations, such as
the leading twist approximation in the OPE and the lowest order
approximation
on the meson--nucleon coupling constant.
Such an approach
allows one to derive a
convolution formula which includes binding effects
and, at the same time,
preserves the
energy--conservation sum rule
{}~\cite{ciofi2}.

In the present paper the model is extended to
deep inelastic scattering of polarized leptons
off polarized nuclear targets.
To this end, the set of operators, providing
the basis for OPE, is extended by
considering the axial operators in terms of  nucleon
fields interacting with meson fields.
Using the
non-relativistic reduction, an explicit  form of the
operators   relevant to describe
polarized deep inelastic scattering are found.
A particular attention is paid to the investigation
of the properties of the physical nucleon, e.g. the
physical mass, the meson cloud, the renormalization constants and the
SSF's
which appear in the calculation.
The explicit expressions for moments of SSF of the lightest nuclei,
the deuteron and $^3He$, are found and
the inverse Mellin transform reconstructs
the corresponding nuclear SSF in
the form of convolution  of the
nucleon SSF
and effective distributions of the nucleons in a nucleus.
The obtained formulae are generalized to the case of heavy nuclei
and found to be similar to
those used in the conventional convolution approach~\cite{ciofi2}.
As an example of applications of the method,
the SSF of the polarized deuteron and $^3He$ are numerically estimated
and compared with recent
experimental data.

The paper is organized as follows. In Section II   the basic
formalism is presented.
The antisymmetric part of the Compton
scattering amplitude
is defined in terms of the axial twist-two operators in the
OPE method and their explicit form
in the non-relativistic limit is
computed. In Section III the moments
of the spin-dependent structure function $g_1^N(x)$ for the physical
nucleon
are evaluated
in terms of the corresponding moments
of bare nucleons and meson cloud contribution.
The moments and the SSF $g_1^{D}(x)$
and $g_1^{^3 He}(x)$ are calculated in Section IV,
where an extension to complex nuclei is proposed and
a formal comparison with the conventional
convolution approach is illustrated. Preliminary results for $A=2$
have already been presented in Ref.\cite{umnm}.
The results of numerical calculations are presented in
Section V and the
Conclusions in Sections VI, respectively.

\section{ Basic formalism}
\setcounter{equation}{0}
\subsection{Kinematics and notation}
The spin-dependent structure function can be determined
experimentally
by measuring the asymmetry in the reaction with polarized particles
$$ \vec l\, + \vec A \longrightarrow\,  l'\,+ \;{\rm X}.$$
In the one-photon exchange approximation the relevant  part of the
cross
section can be written in terms of the antisymmetric parts of the
leptonic $L_{\mu\nu}$  and hadronic $W_{\mu\nu}$  tensors
\begin{equation}
\frac{d^2\sigma}{dE'd\Omega} = \frac{\alpha^2}{Q^4}\frac{E'}{M_A E}
L_{[\mu\nu]}W^{[\mu\nu]},
\label{pg1}
\end{equation}
where $[\mu\nu]$ are the antisymmetric indices, $E$ and $E'$ are the
initial and final energies of the lepton, $Q^2\equiv -q^2$ is the
square  of four-momentum transfer,
$\alpha$ stands for the electromagnetic fine structure constant,
and $M_A$ is the mass of the target.
\widetext
The leptonic tensor is obtained
from
Quantum Electrodynamics and the antisymmetric part
of the hadronic tensor $W_{\mu\nu}$ is expressed
in terms of two independent spin structure functions $G_{1,2}$
\begin{equation}
W_{[\mu\nu]}(p_A, q, S) =
i\epsilon_{\mu\nu\lambda\sigma}\,q^{\lambda}
\left ( G_1(q_0, Q^2)\, S^{\sigma} + \left[ (p_A\cdot q)\, S^\sigma -
(S\cdot q)p^\sigma_A\right] \frac{G_2(q_0, Q^2)}{M_A^2} \right ),
\label{pg2}
\end{equation}
where $p_A$ denotes
the 4-momentum of the target, $q_0$ stands for the time-component of
the
4-vector $q$ (in the rest frame of the target we   use the notation
$q_0\equiv \nu$) and $ S$ is the  polarization four-vector,
normalized as
$S\cdot S = -1$ and satisfying the relation
$S\cdot p_A=0$. In what follows we will consider
deep inelastic scattering off polarized targets with spin one-half
(the nucleon and $^3$He) and
one (the deuteron),
in which case the vector
$S$ may be computed in quantum field theory as the mean value of the
canonical
spin operator defined by Noether's theorem\cite{bog}.\footnote{For
example,
its components are
 $S_\lambda = (0,0,0,\langle \sigma_z \rangle)$
for the nucleon and  $S_\lambda =
-(i/M_D)\epsilon_{\lambda\mu\nu\rho}
{\cal E}^\mu {\cal E}^{\star \nu}
p^\rho_D$, for the deuteron with total momentum $p_D$ and
polarization
4-vector ${\cal E}^\mu$.}
We choose the quantization axis in the opposite direction to
the photon momentum, $q^\mu =(\nu, 0, 0, -\vert {\bf q} \vert)$.

In order to measure the SSF $G_1$ and $G_2$ appearing in eq.
(\ref{pg2}),
one has to consider the
difference between the cross sections
corresponding to parallel and antiparallel electron and target
spins, respectively, for one has
\begin{equation}
\frac{d^2\sigma}{dE'd\Omega}
\left (\uparrow \uparrow\right )-
\frac{d^2\sigma}{dE'd\Omega}
\left (\uparrow \downarrow\right)=
\frac{4\alpha^2 E'}{E Q^2}\left [
\left (E + E'\cos{\theta} \right ) G_1 - Q^2G_2/M_A\right ],
\label{pg3}
\end{equation}
where $\theta$ is the electron
scattering angle. In the Bjorken limit
($  Q^2\to\infty,
\; \nu\to\infty,\; x\equiv Q^2/(2p_Aq)$ is fixed),
which will be considered from now on,
the functions $G_{1,2}$
are predicted to depend only upon $x$, yielding the ``true"
dimensionless spin-dependent structure functions $g_1(x)=\nu
G_1(\nu,Q^2)$ and
$ g_2(x)=\nu^2 G_2(\nu,Q^2)/M_A $.
\narrowtext
Defining
the asymmetry $A_{\|}(x)$ as the ratio
\begin{equation}
A_{\|}(x)=
{{d^2\sigma}/{dE'd\Omega}
\left (\uparrow \uparrow\right )-
{d^2\sigma}/{dE'd\Omega}
\left (\uparrow \downarrow\right)\over
{d^2\sigma}/{dE'd\Omega}
\left (\uparrow \uparrow\right )+
{d^2\sigma}/{dE'd\Omega}
\left (\uparrow \downarrow\right)}~~,
\label{io}
\end{equation}
one obtains
\begin{equation}
A_{\|}(x) = \frac{2xg_1(x)}{F_2(x)},
\label{pg4}
\end{equation}
where $F_2(x)$ is the usual spin-independent structure function.
The asymmetry (\ref{pg4}) and, consequently, the SSF are the main
objects
of experimental  and theoretical  investigations.

Since any structure
function, either spin-independent or spin-dependent, can be
represented
as a linear combinations of
helicity amplitudes, it is instructive to perform
further analysis in the helicity basis. For a thorough analysis of
helicity
 amplitudes in case of targets with an arbitrary spin we refer the
interested reader to~\cite{jaffes}. We define the amplitudes
$A_{\lambda\:{\cal M}, {\lambda^\prime}\:{{\cal M}^\prime}}$ as
\begin{eqnarray}
A_{\lambda\:{\cal M}, {\lambda^\prime}\:{{\cal M}^\prime}} =
{\varepsilon^\mu_\lambda}^{\star}W_{[\mu\nu]}(S)
\varepsilon^\nu_{\lambda^{\prime}},
\label{pg5}
\end{eqnarray}
where $\lambda, \lambda^\prime$ and ${\cal M}, {\cal M}^\prime$
are the spin projections of  the photon and target, respectively,
along the z-axis,
$\varepsilon_\lambda^\mu$ is the polarization vector of a  helicity
$\lambda$
photon, $ \varepsilon_{\pm}^\mu = \mp (0,1,\pm i, 0)/\sqrt{2}, \quad
\varepsilon_{0}^\mu =  (-\vert {\bf q} \vert,0,0,\nu) /
\sqrt{Q^2}\quad $,
and the other notations are self-explaining.
A simple calculation shows that the structure function $g_1$
in the Bjorken limit
is defined by the amplitudes (\ref{pg5})
\begin{eqnarray}
A_{+\pm,+\pm}=\mp g_1\pm\frac{Q^2}{\nu^2}\,g_2
\label{pg5a}
\end{eqnarray}
and it reads
\begin{eqnarray}
g_1 = - \frac{1}{2} \left( A_{++,++} - A_{+-,+-} \right),
\label{pg6}
\end{eqnarray}
Actually, the amplitudes (\ref{pg5}) may contain the  symmetric
part too, which involves the structure functions $F_1(x,Q^2)$ and,
for spin one targets,
$b_1(x,Q^2)$\cite{jaffes}. In our case this part is irrelevant, since
it does not contribute to the asymmetry (\ref{pg4}) and to the SSF.

The SSF,
as well as the hadronic tensor $W_{[\mu\nu]}$, can be directly
calculated, via the optical theorem, from the imaginary part of the
amplitude
for forward Compton scattering of virtual photons off hadronic
targets
\begin{eqnarray}
 W_{[\mu\nu]}=\frac{1}{2\pi} {\rm Im} T_{[\mu\nu]}.
 \label{optica}
\end{eqnarray}
In what follows the Compton amplitude
$T_{[\mu\nu]}$ for nuclear targets and its relation with SSF through
eqs. (\ref{pg5}) and  (\ref{pg6}) will be obtained.

\subsection{The Compton scattering amplitude for nuclear
targets}
To establish a direct connection between SSF and the amplitude
$T_{[\mu\nu]}$ we decompose  the latter in the same form as
(\ref{pg2}),  viz.
\widetext
\begin{eqnarray}
&&T_{[\mu\nu]} =  \epsilon_{\mu \nu \lambda \sigma} q^\lambda \:
\frac{i}{\nu} \: \left(
\alpha_1(x, Q^2)S^\sigma
+ \left(  (p_A\cdot q) S^\sigma -  (S \cdot q)p_A^\sigma \right)
\right)
\frac{\alpha_2(x, Q^2)}{\nu M_A }~~,
\label{pg7}
\end{eqnarray}
\narrowtext
where the Compton spin-dependent structure
functions $\alpha_{1,2}(x)$ are related to the Compton helicity
amplitudes

$h_{\lambda {\cal M},\lambda' {\cal M}'}$ and the deep inelastic SSF
by the
following relations:
\begin{eqnarray}
&&h_{\lambda {\cal M},\lambda' {\cal M}'}= \varepsilon_\lambda^{\mu
\star} T_{[\mu\nu]}(S) \varepsilon^\nu_{\lambda '}
 \label{pg8}\\[1mm]
&&\alpha_1(x) = -\frac{1}{2}(h_{++,++} - h_{+-,+-}); \label{pg8a}\\
&&g_1(x)= \frac{1}{2\pi}{\rm Im} \;\alpha_1(x).
\label{pg9}
\end{eqnarray}
Using  the dispersion relation
for the function $\alpha_1(x)$ and eq. (\ref{pg9}), we get
the expression for $\alpha_1(x)$ in terms of
the moments of the SSF $g_1$

\begin{equation}
\alpha_1(x) = 4\,\sum\limits_{n=0,2,\ldots}^\infty
\left (\frac{1}{x}\right )^{n+1}\int_0^1 dy\,y^n \,g_1(y).
\label{pg15}
\end{equation}
The integral in the r.h.s. in eq. (\ref{pg15}) is  the $n+1$
moment ${\sf M}_{n+1}(Q^2)$ of the SSF. We will use eq.
(\ref{pg15})
to
obtain the structure functions from the
explicit expressions
of the nuclear Compton amplitude $T_{[\mu\nu]}$.

The  computation of the amplitude $T_{[\mu\nu]}$
 requires the treatment of two relatively independent questions:
 i) the analysis of the properties of the
 time ordered product of electromagnetic
current operators at high momentum transfer ($Q^2 \to \infty$),
which characterizes the short distance physics,
and ii) the determination of the vectors of the nuclear ground state
$|p_A\,\rangle$, which  essentially characterizes
the large distance physics.
This is seen explicitly  from the expression of the amplitude
(\ref{pg7}) which
is of the form
\begin{equation}
T_{[\mu\nu]} = i \int d^4 \xi \exp( iq\xi)
\langle p_A \: {\cal M} | T\left ( J_{ \mu }(\xi)J_{ \nu }(0)\right )
| p_A  \: {\cal M}\rangle.
\label{pg11}
\end{equation}
\newcounter{kos}
 \setcounter{kos}{\arabic{equation}}
 \noindent
 The behavior of the $T$-product of the electromagnetic currents may
be
 established in a general form directly from eq.~(\ref{pg11}).
In the Bjorken limit ($Q^2\to\infty$) the main contribution in the
integral (\ref{pg11}) comes from small space-time intervals,
$\,\xi^2\to 0$. In this limit the arguments of electromagnetic
currents
coincide and T-product contains singularities. A consistent
method to analyze these singularities is based
on the Wilson operator product
expansion~\cite{ope}. According
to the OPE on the light-cone~\cite{preparata}, the product
of two arbitrary operators A and B
factorizes into two pieces
when their arguments are separated by
a small space-time interval;
the first piece contains singularities
(the c-number coefficient functions, or Wilson's coefficients) and
the second one
appears as a set of regular local operators,
which are well-defined in field theory.
Then the local operators
are expanded in a  series
\begin{equation}
A(\xi)B(0)\sim \sum_n C_n(\xi^2)\xi_{\mu_1}\ldots\xi_{\mu_n}
O_n^{\mu_1\ldots\mu_n}(0).
\label{pg12}
\end{equation}
The operators $O_n^{\mu_1\ldots\mu_n}$
are defined to be symmetric and traceless
in all Lorentz indices $\mu_1\ldots\mu_n$. The quantity $n$ is
the Lorentz spin. Another quantity, the twist, defined as $\tau = d_n
- n$
($d_n$ - is the canonical dimension of the operator $O_n$)
plays an important
role in the theory of deep inelastic scattering processes.
\widetext
 Namely, only the lowest values of $\tau$ contribute
to the matrix elements of the Compton amplitude~\cite{jaffeb}.
Therefore in the leading
order of the twist ($\tau = 2$), the r.h.s. of eq. (\ref{pg11})
can be  rewritten as
\begin{equation}
{T}_{[\mu\nu]} = i \epsilon_{\mu \nu \lambda \sigma} q^\lambda
\sum_{t;n=0,2,\ldots}^{\infty}
C_{n,t}(Q^2)
{\left( \frac{2}{Q^2}\right)}^{n+1} q_{\mu_1} \ldots q_{\mu_n}
\langle \, p_A\,{\cal M}|
{\hat{O}}_t^{\{\sigma \mu_1 \ldots \mu_n\}}(0)|p_A\, {\cal
M}\,\rangle,
\label{pg13}
\end{equation}
where  $t$ tags the fundamental fields of the theory under
consideration
and ${\hat{O}}_t^{\{\sigma \mu_1 \ldots \mu_n\}}(0)$ are the relevant
twist-two
operators constructed from this fields.
The transformation properties of the amplitude $T_{[\mu\nu]}$
 restrict the Lorentz spin, ($n+1$), in eq. (\ref{pg13}) to take only
odd values.

It is worth
emphasizing that in (\ref{pg13}), due to the factorization in the
OPE,
the coefficient functions $C_n$ are
related to short distance, ``subhadronic'', physics (depending on
the large momentum $q$), whereas the matrix
elements of the operators $O^{\{\sigma \mu_1\ldots\mu_n\}}$
characterize the
large distance physics (depending on typical nuclear  momenta).

In order to separate  the part contributing  to
$g_1(x)$, it is convenient to rearrange the symmetric Lorentz indices
$\{\sigma\mu_1\ldots\mu_n\}$ so as to obtain two operators,
one with no definite
symmetry ${\hat{O}}^{\sigma \{\mu_1 \ldots \mu_n\} } $, and the other
with
mixed symmetry~\cite{manohar},  viz.
\begin{eqnarray}
{\hat{O}}^{\{\sigma \mu_1 \ldots \mu_n\}} =
\left( {\hat{O}}^{\sigma \{\mu_1 \ldots \mu_n\}} +
\frac{1}{n+1} \left[
\sum_{i=1}^{n}
{\hat{O}}^{\mu_i \{\mu_1 \ldots \sigma \mu_{i+1} \ldots \mu_n\}} -
n \cdot {\hat{O}} ^{\sigma \{\mu_1 \ldots \mu_n\}}\right] \right).
\label{pg14}
\end{eqnarray}
\narrowtext
 Then $g_1$ gets contributions from
${\hat{O}}^{\sigma \{\mu_1 \ldots \mu_n\} }$ only. The
mixed symmetry operators
enclosed in
the square
brackets of eq. (\ref{pg14}), define the twist-2 part of $g_2(x)$,
the so--called Wandzura-Wilczek contribution~\cite{manohar,ww}.
 In this paper we consider the amplitude $T_{[\mu\nu]}$ defined by
 eqs.~(\ref{pg7}) and (\ref{pg13}) with only the first operator
from eq. (\ref{pg14}), that is, the part concerning the
structure function $g_1(x)$ only. The structure function
$g_2(x)$  is expected to be very small,
(in the parton model it is exactly zero),
 and will not be considered here (for the investigation of nuclear
effects
 in $g_2(x)$ see, for instance Ref.~\cite{mankiewicz}).

The form  (\ref{pg13}) for the amplitude (\ref{pg11})
  near the light-cone is valid
  in the framework of any renormalizable field
theory~\cite{ope,preparata}.
  Thus, the problem of its analysis is formulated now
  as a consistent calculation of both the coefficient
  functions $C_{n,t}$ and the matrix elements of the twist-two
operators
  $O^{\sigma\{\mu_1\ldots\mu_n\}}$ sandwiched between nuclear ground
  state vectors $|p_A\rangle$. So far there does not exist a
realistic
  field theory by which one could compute simultaneously both pieces,
since
  if one of them is calculated in a more or less
  self-consistent way, then
the other one relies on phenomenological approaches.

For example, within perturbative QCD, because of asymptotic freedom
  it is possible to analyze the properties of the Wilson's
coefficients
  $C_{n,t}(\alpha_s,Q^2)$
  by computing directly, at least in principle, the corresponding
Feynman graphs
  up to the desired order in $\alpha_s $~\cite{muta}.

  Making use of the renormalization
  group equations, the $Q^2-$dependence of these coefficients
  may be  established as well. However, in
  practical calculations of the structure functions and comparison
with
  experimental data one needs the  matrix elements of the
  operators $\hat O_t(0)$, which are related to the
  non-perturbative region of QCD and, hence, are parameterized from
the
  experimental data.
  Since the basis of operators in OPE, eq. (\ref{pg13}),
  and the Wilson coefficients
  are target independent, all information about
  the target is contained
  in the unknown matrix elements.
  Therefore in the treatment of the ``QCD motivated" models of
nuclear
  effects in
  deep inelastic scattering,
one is forced to introduce
  parameters, which neither can be computed
  theoretically from QCD,
  nor can be fixed from independent experiments~\cite{rescaling}.

The problems related to the short and
  large distance are formally solved, in case of deep inelastic
  scattering, within
  non-asymptotically free theories with spinor fields (nucleons)
  interacting with the massive bosons (mesons)
  via pseudoscalar and vector
  couplings~\cite{gribov,m-n nauka}.
  These field theoretical models
  with a renormalizable interaction allow
  a perturbative investigation of the Wilson's coefficients and of
  the corresponding matrix elements, by summing the leading
logarithmic
corrections.
  These examples can be considered as an
  idealized meson-nucleon
  theory, since the realistic models necessarily involve
  phenomenological corrections, such as vertex meson-nucleon form
  factors, effective coupling constants and meson
  masses~\cite{brown,weise}.
  In realistic meson-nucleon models, the exact results of
  Refs.~\cite{gribov,m-n nauka} serve as a hint for a formal
approach and further phenomenological
  adjustments.
  It is obvious that
  the target independent Wilson's coefficients are not calculable
 within such a theory and they ought to be parameterized from
experiments, for instance, from the experiments on deep inelastic
  scattering off  free nucleons.

  We apply the effective meson-nucleon theory  and the OPE method
  to deep inelastic scattering off nuclei (c.f.
  Refs.~\cite{levin,kaz,ukhanna,mankiewicz}).
  This theory allows one to describe fairly  well the NN-interaction
  at relatively
  small energies, the nuclear bound states
  $|p_A\,\rangle$, the binding energy and other properties of light
nuclei
{}~\cite{bonn}.
 Then, the axial twist-two operator $O^{\sigma\{\mu_1\ldots\mu_n\}}$
for the
 system of interacting nucleon and meson fields
 may be written down in the form
\begin{eqnarray}
{\hat{O}}_n^{\sigma\{ \mu_1 \ldots \mu_n\}}(0) &=&
{\left(\frac{i}{2}\right)}^n \nonumber \\
&\times&
 \{ :\!\bar{N}(0) \gamma^\sigma \gamma_5
\stackrel{\leftrightarrow}{\partial^{\mu_1}} \ldots
\stackrel{\leftrightarrow}{\partial^{\mu_n}} N(0)\!: \},
\label{pg16}
\end{eqnarray}
where $N$ stands for the  nucleon spinor fields.
Note the implicit presence of the  meson degrees of freedom,
 via interaction, in the definition of
the operator ${\hat{O}}_n^{\sigma\{ \mu_1 \ldots \mu_n\}}(0)$.
In order to obtain the explicit expressions of the operator
(\ref{pg16}) and calculate its matrix elements, we need the
Hamiltonian of the system. In order to achieve
self-consistency, this
Hamiltonian has to provide simultaneously the equation of motion
for the interacting fields and the target ground state:
\begin{eqnarray}
& &
\dot N = i\left [ H,N \right ],
\label{pg170}\\[1mm]
& & H \, |p_A \rangle = M_A | p_A \rangle .
\label{pg17}
\end{eqnarray}
Equation (\ref{pg17}) for the nuclear
ground state has been solved
within the non-relativistic limit
using effective Hamiltonians
containing $\pi,\, \sigma,\, \omega,\, \rho,\, \eta$ and $\delta $
mesons
(one-boson-exchange
(OBE) approximation)~\cite{brown,bonn}.
Consequently, the operators (\ref{pg16}) are also to be
calculated within the non-relativistic limit.
The strategy of our calculation
is therefore as follows:
  i) we choose the appropriate covariant Lagrangian, giving
the classical equation of motion for interacting meson and
nucleon fields;
 ii) these  equations are reduced non-relativistically
 and the effective non-relativistic Hamiltonian
is obtained;
iii) using the same procedure of non-relativistic reduction,
we compute the explicit form of the operators (\ref{pg16}).

The procedure of non-relativistic reduction of classical equations
of motion for the interacting meson and nucleon fields has been
established in a number of papers and could be found in details,
for instance, in Refs.\cite{rosaclot,kunucl}.

Below we perform explicitly the
calculations with pseudoscalar-isovector coupling, the pion-nucleon
interaction. The result is generalized  to the
case of other kinds of couplings.
Introducing the
isospin formalism
we redefine the Wilson's coefficients in eq.(\ref{pg13})
as a diagonal $(2\times 2)$ matrix  in the isospin space with the
proton and neutron coefficients on the main diagonal,
$\left ( \hat C_n \right )_{\alpha\beta}=
C_{n,\alpha}\delta_{\alpha\beta}$,
 $\alpha,\,\beta =1,2$,
 and use them into the definition of the operators (\ref{pg16}).

\widetext
The contribution of the operators ${\hat{O}}^{\sigma \{\mu_1 \ldots
\mu_n\} }$
(\ref{pg16}) to the Compton helicity amplitude
then becomes
\begin{eqnarray}
h_{+ \: {\cal M},+ \: {\cal M}}&=&
 - \sum_{n=0,2,\ldots}^{\infty}
{\left( \frac{2\nu}{Q^2}\right)}^{n+1} \: {\langle O^+_n \rangle}_A,
\nonumber\\[1mm]
{\langle O^+_{n}  \rangle}_A &=&
{\left( \frac{i}{2}\right)}^{n} \langle p_A\: {\cal M} |
:\!\bar{N }(0) \hat C_n(Q^2)\,
\gamma^{+} \gamma_5 {
\stackrel{\leftrightarrow}{\partial_{-}}}^n N (0)\!: | p_A\: {\cal M}
\rangle ,
\label{pg19}
\end{eqnarray}
where $\gamma_+ = \gamma_0 + \gamma_z$ and $\partial_- =
\partial_0 - \partial_z$.

We perform the non-relativistic transition of the fields $N(0)$,
following the method described in Refs.\cite{rosaclot,kunucl},
and, using the equation of motion (\ref{pg17}), we compute their n-th
$\partial_- $ derivatives.
The resulting operators, being composed from interacting nucleon
fields,
explicitly involve
the meson degrees of freedom.
Skipping some rather cumbersome
details of calculations we write below
 the explicit non-relativistic form of the operators $O^+_{n} $
as a sum of operators up to second order of growth coupling constant
$g_\pi$,
up to $g^2_\pi$:
\begin{eqnarray}
O^\lambda_{n}  = O^\lambda_{N,n}  + O^\lambda_{N \pi,n}  +
 O^\lambda_{N \: N,n},
\label{pg20}
\end{eqnarray}
\begin{eqnarray}
O^\lambda_{N,n}&=& m^n \int \frac{d{\bf p}_1 d{\bf p}_2 }{
{(2\pi)}^6}
 {\sl N}_n^{(1)}({\bf p}_1, {\bf p}_2)
{\left[ \Sigma^\lambda ({\bf p}_1, {\bf p}_2) \right] }_{{s_1} \:
{s_2}}
a^{+} ({\bf p}_1, {s_1}) \,\hat C_n\, a ({\bf p}_2, {s_2}),
\label{pg21}
\\[2mm]
O^\lambda_{N\pi,n}&=&m^n\frac{i g_\pi}{2m}
\int \frac{d{\bf p}_1 d{\bf p}_2}{{(2\pi)}^9}
\frac{d{\bf k}}{\sqrt{2\omega({\bf k})}}
\left(
{\sl N}_n^{(-)}({\bf k}) \, b_j^{+}({\bf k}) -
{\sl N}_n^{(+)}({\bf k}) \, b_j ({\bf k})
\right) \nonumber \\
&\times&
{\left[
\Sigma^\lambda ({\bf p}_1, {\bf p}_2) \, {\bfgr \sigma} \cdot {\bf k
}
\right]}_{{s_1} \: {s_2}} \;
a^{+} ({\bf p}_1, {s_1}) \hat C_n\, {\tau^j}  \, a ({\bf p}_2, {s_2})
+ h.c.,
 \label{pg22} \\[2mm]
O^\lambda_{NN,n}&=&- m^n \frac{{g_\pi}^2}{4m^2}
\int \frac{d{\bf p}_1 d{\bf p}_2 d{\bf p}}{{(2\pi)}^{12}}
\frac{d{\bf k}}{2\omega({\bf k})}
{\left[
\Sigma^\lambda ({\bf p}_1, {\bf p}_2) \, {\bfgr \sigma} \cdot {\bf k
}
\right]}_{{s_1} \: {s_2}}
\left({\sl \tilde{N}}_n^{(-)}({\bf k}) +
{\sl \tilde{N}}_n^{(+)}({\bf - k}) \right)\nonumber \\
& &
{\left[{{\bfgr \sigma} \cdot  {\bf k }}\right]}_{s_3 \: s_4}
:a^{+} ({\bf p}_1, {s_1}) \hat C_n \tau^j\,a ({\bf p}_2, {s_2})
 a^{+} ({\bf p}, {s_3}) \,\tau_j\,a ({\bf p+k}, {s_4}):
 +h.c.
\label{pg23}
\end{eqnarray}
where $m$ is the  mass parameter of the Lagrangian, the mass
of bare nucleons,
$\tau_j$ are the isospin matrices,
$a^\pm ({\bf p},s) $ are  creation and annihilation operators
of a bare nucleon, proton or neutron,   with spin $s$ and momentum
${\bf p}$, $b^{\pm}_j({\bf k})$ create or annihilate the $j$-th pion
with momentum {\bf k};
$\omega = \sqrt {m_\pi^2 + {\bf k}^2}, \quad$
($m_\pi$ is the pion mass),
$ {\left[ \Sigma^\lambda \right]}_{s_1 \: s_2} = \chi_{s_1}^+ \:
\Sigma^\lambda \: \chi_{s_2}$; $\chi_s$ is the Pauli spinor
and the operator $\Sigma$ is   the non-relativistic
analogue of the four-vector spin operator for the spin-$\frac{1}{2}$
particles
\begin{eqnarray}
\Sigma^\lambda({\bf p}_1,{\bf p}_2) = \left\{
     \begin{array}{ll}
     \frac{1}{2m} {\bfgr \sigma} \cdot
\left( {\bf p}_1 + {\bf p}_2 \right),   &  \; \lambda = 0 \\[2mm]
 \sigma^\lambda \left(1 - \frac{{{\bf p}_1}^2  +
{{\bf p}_2}^2}{8m^2}\right) + & \phantom{...}  \\[2mm]
     + \: \frac{1}{4m^2}
     \left(
     {\bfgr \sigma} \cdot {\bf p}_1 \: p_2^\lambda +
     {\bfgr \sigma} \cdot {\bf p}_2 \: p_1^\lambda -
 {\bf p}_1 \cdot {\bf p}_2 \: \sigma^\lambda - i {[{{\bf p}_1 \times
     {\bf p}_2}]}^\lambda \right),
     &  \; \lambda = x, y, z
     \end{array} \right.
\label{pg24}
\end{eqnarray}
The functions $N_n^{i}$ in eqs. (\ref{pg21})-(\ref{pg23}) depend only
upon momenta and are of the form
\begin{eqnarray}
&&{\sl N}_n^{(1)}({\bf p}_1,{\bf p}_2)= {\left(
1 + \frac{{{\bf p}_1}^2 + {{\bf p}_2}^2}{4m^2} + \frac{{p_1}_z +
{p_2}_z}{2m}
\right)}^n,
\label{pg25} \\[2mm]
&&{\sl N}_n^{(\pm)}({\bf k})= \pm \; \frac{1}{{\omega_+({\bf k})}}
\left( {\left[ 1 \pm \frac{{\omega_+
({\bf k})}}{2m} \right]}^n - 1 \right),
\label{pg26}\\[2mm]
&&{\sl \tilde{N}}_n^{(\pm)}({\bf k}) = \frac{1}{\omega({\bf k})}
{\sl N}_n^{(\pm)}({\bf k}) \mp \; \frac{1}{\omega({\bf k})}
\frac{\omega_+({\bf k})} {k_z^2}
\left( {\left[ 1 \pm \frac{k_z}{2m} \right]}^n - 1 \right ),
\label{pg27}
\end{eqnarray}
where  $\omega_{+}\equiv\omega +k_z$.
The operators (\ref{pg21})-(\ref{pg23}) are the basic result of
non-relativistic OPE method within the OBE approximation. Their
matrix elements
will determine the polarized deep inelastic scattering
of leptons on those nuclear targets,
e.g. the two and three nucleon systems, which are well described
within
the effective meson-nucleon theory with OBE potential.

\narrowtext
  Note that our method is based upon perturbation theory
  within the effective meson-nucleon theory.
  Since the effective meson-nucleon coupling constants are large, it

can be
  argued whether a perturbation expansion can be applied at all.
We simply follow here the
customary use of perturbation
  theory in the effective   meson-nucleon
theory~\cite{brown,weise},
justified by the success of OBE model.
The price one has to pay in such an approach is the use of
some phenomenological ingredients. For example,
  since the nuclear forces are strongly repulsive at small distances,
  the physics at such distances is
mocked up by a
  short-range
  repulsive core, which is handled
  partly by introducing
  meson-nucleon vertex formfactors.
  Thus, in the calculation
  of the  corresponding matrix elements the uncertainties of the
nuclear
  force at very short distances are insignificant.  Analogously the
  short distance contributions to the nucleon and nuclear structure
  functions
  are hidden into the Wilson's coefficients, which are the new
``effective
  constants'' in the effective meson-nucleon
  theory~\cite{kaz,ukhanna,mankiewicz,levin}.

  From the QCD point of view this situation corresponds to the
picture
  when hadrons, viewed as
  a core of valence quarks surrounded by
  a sea quark-antiquark pairs and gluons, are approximated by
  a core of  bare nucleons with a correlated color neutral
  quark-antiquark pairs, the meson cloud. In OPE such a picture
  means that   much of detailed dynamics of the quarks is embedded
  in the coefficient functions $C_n(Q^2)$, the influence of the meson
  cloud is rather included into the matrix elements of the operators
  $O^{\sigma\{\mu_1\ldots\mu_n\}}$.  In case of the effective
 meson-nucleon theory a preliminary investigation of the twist
 behavior of eq.~(\ref{pg13}) is hindered by the fact that all
 the renormalization effects are included into the effective
constants
 of the theory (coupling constants, meson masses, vertex formfactors
etc.),
 so that the renormalization group equations are here, in a sense,
 inapplicable. The use of the twist-two as a leading term in
 eq.~(\ref{pg13}) is to be regarded as an assumption and it should be
 verified ``a posteriori'', by looking at the quality of
the final results and,
  possibly,
 by investigating the higher twist corrections.

\section{Moments of the  nucleon structure functions}
\setcounter{equation}{0}
As  mentioned above, the non-relativistic expressions for the
 operators (\ref{pg21})-(\ref{pg23}) and the Compton amplitude
(\ref{pg19}) involve, in particular,  the  bare (unknown) parameters,
i.e.
the mass parameter from the Lagrangian of the theory and the
Wilson's coefficients $C_n(Q^2)$ which are not calculable within our
approach. Whereas the
mass parameter may be fixed by introducing into the
Lagrangian a corresponding counterterm, the coefficients $C_n(Q^2)$
are to be related with the SSF of the physical nucleons. Therefore we
first investigate the ground state and the SSF of the physical
nucleons.

\subsection{Nucleon ground state}
For the investigation of the nuclear ground states in the effective
 meson-nucleon theory it is convenient to use the Tamm-Dancoff
method.
\widetext
For a physical nucleon with an isospin index $\alpha$,
 momentum ${\bf q}$ and a given z-projection of the spin $s$,
the Tamm-Dancoff decomposition is given by
\begin{eqnarray}
| N \rangle_{{\bf q},\: \alpha, s} = \sqrt{1-Z_N} \varphi_0 | \bar{N}
\rangle_{{\bf q},\: \alpha, s}
+ \varphi_1 | \bar{N} \pi \rangle_{{\bf q},\: \alpha, s} + \ldots,
\label{pg28}
\end{eqnarray}
where $Z_N$ is the normalization constant defined by the condition
$\langle N|N \rangle =1$ and $|\bar {N}\rangle$ and
$|\bar {N}\pi\rangle$ represent
the basis vectors of
the states with one bare nucleon,
and with one bare nucleon and one meson,
respectively. The coefficients $\varphi_i$
in the expansion (\ref{pg28}) are the  operators
in the momentum space and define the corresponding wave functions
\begin{eqnarray}
 \varphi_0 | \bar{N} \rangle_{{\bf q},\:\alpha, s}&=&a^+_\alpha({\bf
q}, s) |0\rangle, \label{pg290} \\
 \varphi_1 | \bar{N} \pi \rangle_{{\bf q},\: \alpha, s}&=& \int
\frac{d{\bf k}d {\bf p}}{{(2\pi)}^6}
 \varphi^{\alpha^\prime, s^\prime, \: j}_{1 s, \: \alpha}({\bf p},
{\bf k},
{\bf q}) a^+_{\alpha^\prime}({\bf p}, s^\prime) b^+_j({\bf k}) | 0
\rangle.
\label{pg29}
\end{eqnarray}

The explicit expressions for the wave functions $\varphi_1$
is found from the condition that the
$|N\rangle$ is the state of the
physical nucleon ,
which means  that it obeys
eq. (\ref{pg17}) with the physical nucleon mass\footnote{In eq.
(\ref{pg28}) we keep only the first two terms. The next term,
$\varphi_2 |\bar N\pi\pi\rangle$, in spite
of being proportional to $g^2_\pi$, does not contribute to the
nucleon
SSF, as it can be seen from eqs. (\ref{pg21})-(\ref{pg29}).}
, viz
\begin{eqnarray}
\varphi^{\alpha^\prime \:, s^\prime,\: j}_{1 M, \: \alpha}({\bf p},
{\bf k}, {\bf q}) = - \:
{(2\pi)}^3 \delta^{(3)}\left( {\bf p + k  - q} \right) \: \frac{i
g_\pi}{2m_N} \:
\frac{ {\left[{\bfgr \sigma} \cdot {\bf k} \right]}_{s^\prime \: M} }
{ \omega({\bf k}) \sqrt{ 2 \omega({\bf k}) } }
\: { \left[ \tau^j  \right] }_{\alpha^\prime \: \alpha}.
\label{pg30}
\end{eqnarray}
In particular, as it can be seen from (\ref{pg30}),
in the non-relativistic approximation
there is no interaction between
the nucleon and the pion with the angular orbital momentum $l \neq
1$.
Finally one can find the mass counterterm
 $\delta m$ and the renormalization constant $Z_N$  that are of the
form
\begin{eqnarray}
\delta m =  - \frac{g^2_\pi}{4m^2}
\int \frac{d {\bf k}}{{(2\pi)}^3} \frac{3 {\bf k}^2}{2 \omega^2({\bf
k})},
\quad \quad
Z_N =  \frac{g^2_\pi}{4m^2} \int
\frac{d {\bf k}}{{(2\pi)}^3} \frac{3 {\bf k}^2}{2 \omega^3({\bf k})}.
\label{pg31}
\end{eqnarray}

\subsection{The matrix elements}
Now we are in the position to compute the matrix elements of the
operators
(\ref{pg21}) and (\ref{pg22}) with
nucleon ground state vectors (\ref{pg28}).
 The operator (\ref{pg23}) is of two-body origin, hence it does not
contribute to the nucleon matrix elements.

Schematically the matrix elements for a nucleon at rest are as
follows
\begin{eqnarray}
{\rm M.E.} &\sim&
_{\alpha,s_z}\langle\bar N|\varphi_0\sum_{n} O_{N,n}^+
\varphi_0 |\bar N\rangle_{\alpha,s_z} \,(1-Z_N)
\quad\quad\quad {\rm (IA + renorm.)}\label{pg32}\\[1mm]
&+&
_{\alpha,s_z}\langle\bar N,\pi|\varphi_1\sum_{n} O_{N,n}^+
\varphi_1 |\bar N,\pi\rangle_{\alpha,s_z}
\quad\quad\quad \quad\quad {\rm (recoil)}\label{pg33}\\[1mm]
&+&
_{\alpha,s_z}\langle\bar N,\pi|\varphi_1\sum_{n } O_{N\pi,n}^+
\varphi_0 |\bar N\rangle_{\alpha,s_z} +\, {\rm h.c.}
\quad\quad\quad {\rm (interaction)}\label{pg34}
\end{eqnarray}
The  four different matrix elements given by eqs.
(\ref{pg32})-(\ref{pg34}) are known, respectively,
as the impulse approximation
(scattering off bare constituents), the renormalization and recoil
contributions, and the term of pure interaction origin
(self energy--like correction). The contribution
of these matrix elements  with  given isospin
to the helicity amplitude
 (\ref{pg19}), sandwiched between the states
$\alpha$, may be explicitly written in the form
\begin{eqnarray}
&-&\frac{1}{2}h_{++\,,++}^{\alpha\alpha}=
  \sum_{n=0,2,\ldots}^{\infty}
{\left( \frac{1}{x}\right)}^{n+1}
 \left [\left(  \hat C_n    \right)_{\alpha\alpha}
\left\{
{\left(1 - \frac{\delta m}{m_N} \right)}^{n} \! \! - Z_N \right \}
\right . \label{pg35} \\[1mm]
&+&\left( \tau^j\,\hat C_n \,\tau_j \right )_{\alpha\alpha}
\frac{g^2_\pi}{4m_N^2}
\left\{
\int \frac{d{\bf k}}{(2\pi)^3}
\frac{2k_z^2-{\bf k}^2}{ \omega^2({\bf k})}
 \left ( \frac{1}{2\omega({\bf k})} -
{\sl N}_n^{(-)}({\bf k}) \right ) \right \}  \label{pg36}\\[2mm]
&-&  \left .
\left(  \hat C_n  \right)_{\alpha\alpha}
\frac{g^2_\pi}{4m_N^2}
 \int \frac{d{\bf k}}{{(2\pi)}^3}
{\sl N}_n^{(+)}({\bf k})
\frac{ 3 \: {\bf k}^2}{\omega^2({\bf k})} \right ],
\label{pg37}
\end{eqnarray}
where $\alpha=1,2$ corresponds to the matrix elements on proton and
neutron respectively.

\narrowtext
The physical meaning of the obtained result becomes clear if
eqs. (\ref{pg35})-(\ref{pg37}) are depicted in terms of Feynman
diagrams.
Fig.~1 represents the helicity amplitude for the proton.
It can be seen that the proton amplitude is determined not only by
the proton Wilson coefficients $C_{n, 1 }$, but by
the neutron coefficients  $C_{n, 2 }$ as well
multiplied by a
factor 2 which comes from  the  relation between the pion-nucleon
coupling constants: $g_{\pi^\pm N} = \sqrt{2}\,g_{\pi^0 N}$.
{}From eqs. (\ref{pg9}) and (\ref{pg15}) it is easy to show that
in the present approach the Wilson's coefficients in the OPE
are proportional to the moments of SSF of the bare nucleons
(proton (p) on neutron (n))
\begin{eqnarray}
C_{n,1(2) } =  2 \, { \bar{\sf M}}_{n+1}
({\bar g}_1^{p(n)}),
\label{pg37a}
\end{eqnarray}
with
\begin{eqnarray}
{\sf M}_n(f) \equiv \int \limits_0^1 dx \: x^{n-1} f(x).
\label{pg38}
\end{eqnarray}
Relations (\ref{pg35})-(\ref{pg38}) determine the
moments of SSF of the physical nucleons
in the effective meson-nucleon theory. Accordingly,
the moments should be
parameterized in order
to describe the scattering off free physical nucleons, and
they should be replaced by this parameterization
whenever they appear in nuclear matrix elements.
The
first ($n=0$) moments of the nucleon SSF play an
important role in deep inelastic scattering of polarized particles,
since
they define
some integral relations among the SSF, known as sum rules.
As $n=0$ the interaction term in eqs.~(\ref{pg36})-(\ref{pg37})
vanishes
(see also eqs. (\ref{pg24})-(\ref{pg26})) and one obtains
\begin{eqnarray}
{\sf M}^{p(n)}_1(g_1) &=&\bar{\sf M}_1^{p(n)}\, (1-Z_N)
\nonumber \\
&+&
\left ( \bar{\sf M}^{p(n)}_1 + 2\bar {\sf M}^{n(p)}_1\right )
\nonumber \\
&\times&
\frac{g^2_\pi}{4m_N^2}
\int \frac{d{\bf k}}{{(2\pi)}^3}
\frac{2k_z^2-{\bf k}^2}{ \omega^2({\bf k})}
\label{pg39}
\end{eqnarray}
The second term in r.h.s. of eq. (\ref{pg39}) is the contribution
of recoil diagrams, which is partially canceled by the
renormalization term~\cite{mec}.
The remaining part is the additional counterterm  to renormalize
the composite axial operator (\ref{pg16}).
For $n=0$ the  operator  (\ref{pg16})
 is the spin operator for the spinor  fields
and its matrix element, (\ref{pg39}), is the mean value
of the spin projection  over polarized states
of the physical nucleon.
Equation (\ref{pg39}) demonstrates that the spin projection of the
physical
nucleon {\em differs} from the mean value of the
spin projection of the  bare nucleons. This is expected, since the
interaction
between the core and the meson cloud, with orbital momentum $l=1$,
slightly redistributes the spin among the constituents.
The total angular
momentum of a meson-nucleon system, is a sum of the orbital
momentum $l=1$ and the spin $s={1}/{2}$, so that both
parallel and antiparallel polarizations of the core
contribute to the nucleon polarization.
 From the point of view of the present approach, the
orbital moment of the meson cloud affects the spin distribution of
bare nucleons inside a  polarized physical nucleon.
The  effects of the meson cloud on
unpolarized deep inelastic scattering
on nucleon have been investigated in refs.\cite{sull,sull1}
and are known as the Sullivan processes.

Another interesting  consequence of the eq. (\ref{pg39}) may be
derived
by analyzing the difference between the first moments
of SSF of the proton and neutron, the Bjorken sum rule.
In this case  one finds that the Bjorken sum rule on free nucleons
may be computed by considering the bare constituents
and the meson cloud corrections

\begin{eqnarray}
\int \left ( g_1^p(x,Q^2)\right . &-& \left .
g_1^n(x,Q^2) \right ) dx \equiv
{\sf M}_1^{p}(Q^2) - {\sf M}_1^{n}(Q^2) = \nonumber \\
&=&
\left ( \bar{\sf M}_1^{p}(Q^2) - \bar{\sf M}_1^{n}(Q^2)\right )
\nonumber \\
&\times&
\left ( 1-\frac{g_\pi^2}{3m^2_N}\int \frac{d {\bf k} }{(2\pi)^3}
\frac{{\bf k}^2}{\omega^3({\bf k})}\, \right ),
\label{pg40}
\end{eqnarray}
where the r.h.s. of eq. (\ref{pg40}) is the BSR for bare nucleons
corrected by the interaction with the meson cloud.
Explicit numerical estimates of the role of the meson
cloud in polarized deep inelastic scattering processes on nucleons
is ambiguous and requires a proper investigation. In fact
one needs
a consistent procedure for the regularization of integrals appearing
in eq. (\ref{pg40}) and for the inclusion of the vertex form factor
into the meson-nucleon vertex. Also, some assumptions about the
properties of the bare nucleon core are necessary. The role
of the orbital momentum of the virtual pions in determining
the spin of the nucleon has to be investigated as well.
These problems
are beyond the goal of the present paper and will be considered
elsewhere.

\section{Nuclear matrix elements}
\setcounter{equation}{0}
In the previous section we explained in details the basic
formalism and apply it to the physical nucleons. The basic idea of
the
calculations of nuclear matrix elements remains the same:
once the non-relativistic expressions for the axial operators
have been established in terms of nucleon and meson fields,
the vectors of nuclear ground states are to be defined in the same
manner, i.e., making use of the same  non-relativistic Hamiltonian
which
has been used to derive eqs. (\ref{pg21})-(\ref{pg29}).
In what follows we keep the terms up to
the second order in the meson-nucleon coupling constant $g$, which
corresponds to the usual approximations in nuclear physics in
deriving the potential and Schr\"odinger equation,
$g^2$-approximation.
In this sense the present approach
pretends to be self-consistent. The condition of a  consistency
is that the Hamiltonian, with OBE interaction, should indeed describe
the real nucleus. We choose the realistic OBE potentials, such as the
Bonn \cite{bonn} or Reid \cite{reid} ones, which
give a good description of light nuclei.

Below the explicit expressions for the moments of the $^2H$
and $^3He$ SSF's, are
derived.

\subsection{The deuteron}
\widetext
The vectors of the deuteron ground state in the Tamm-Dancoff
approximation
are given by a relation
similar to eq. (\ref{pg28}); viz
 \begin{eqnarray}
&&
\varphi_0\,|\bar N\bar N> \, =
\varphi_{0 }^{s_1s_2 }\,({\bf p_1},{\bf p_2})\,
(2\pi)^3\delta ({\bf p_1}+{\bf p_2})\, a^+_{\alpha}({\bf p_1},s_1)
\frac{\epsilon^{\alpha\beta}}{\sqrt{2}}
 a^+_{\beta}({\bf p_2},s_2)\,|0> \label{pg41}\\[1mm]
&&\varphi_1\,|\bar N\bar N\pi> \, =\nonumber\\[1mm]
&&\varphi_{1 }^{s_1s_2\alpha\beta  j}\,({\bf p_1},{\bf p_2},{\bf
k})\,
(2\pi)^3\delta ({\bf p_1}+{\bf p_2}+{\bf k})\, a^+_{\alpha}({\bf
p_1},s_1)
 a^+_{\beta}({\bf p_2},s_2)b^+_j({\bf k})\,|0> \label{pg42},
\end{eqnarray}
where $\epsilon^{\alpha\beta}$ is the Levi-Civita tensor which
describes
the isospin function of the deuteron and the nucleon spins
$s_1,\,s_2  $ and orbital momentum
in $\varphi_{0 }^{s_1s_2 }\,({\bf p_1,  p_2})\,$
are summed to the total angular momentum $J=1$. The quantum numbers
$\alpha $, $\beta $ and $j$ are
combined so as to give the total deuteron isospin $T=0$.
The physical meaning of the coefficient $\varphi_0$ can be understood
by projecting the bare deuteron state onto the state
with two nucleons located at points
${\bf r_1},\,{\bf r_2}$ in co-ordinate space
\begin{eqnarray}
&&< {\bf r}_1,\,{\bf r}_2|\varphi_0\bar N_1\bar N_2>\nonumber\\[1mm]
&&= \int\frac{ d {\bf k}_1 d {\bf k}_2}
{(2\pi)^6} {\rm e}^{i{\bf k}_1 {\bf r}_1+i{\bf k}_2{\bf r}_2}
(2\pi)^3\delta ({\bf p_1}+{\bf p_2})\,
\varphi_{0 }^{s_1s_2}\,({\bf p_1},{\bf p_2})\,
 \chi_{\tilde s_1}
 \chi_{\tilde s_2} \eta_{\tilde \alpha} \eta_{\tilde\beta}
\frac{\epsilon^{\alpha\beta}}{\sqrt{2}} \nonumber \\[1mm]
&\times&
\langle\,0|a_{\tilde \alpha}({\bf k_1},\tilde s_1)
 a_{\tilde\beta}({\bf k_2},\tilde s_2)
a^+_{\alpha}({\bf p_1},s_1)
 a^+_{\beta}({\bf p_2},s_2)|0\,\rangle,
\label{pg43}
\end{eqnarray}

It is clear from eq. (\ref{pg43}) that
$\varphi_0({\bf p, p})$  with spin,
 $\chi_{s}$, and isospin, $\eta_\alpha$,
functions is the conventional deuteron wave function
in the momentum space, obeying
the Schr\"odinger equation with OBE potential~\cite{rosaclot}.
In what follows a more conventional notation
for
$\varphi_0$ will
be adopted,  viz.
$\varphi_0 ({\bf p,p})\equiv \Psi_{\cal M}^D({\bf p})$.
For the function $\varphi_1$ we have
\begin{eqnarray}
&&\varphi^{s_1s_2\alpha\: \beta,j}_{1 }({\bf p}_1,{\bf p}_2,{\bf k})
 =
- {(2\pi)}^3 \delta^{(3)}\left( {\bf p}_1+{\bf p}_2 + {\bf k} \right)
\:
\frac{i g_\pi}{2m_N} \:
\frac{1}{ \omega({\bf k}) \sqrt{ 2 \omega({\bf k}) } }
\nonumber\\[1mm]
&\times&
\left\{
\varphi_{0 }^{s_1s }\,({\bf p}_1,{\bf p}_1)
\frac{{\epsilon}^{\alpha\alpha'}}{2}
\left[ {\bfgr \sigma} \cdot {\bf k} \right]_{s \: s_2}
\left[ \tau^j \right]_{\alpha'\beta}
-(1\leftrightarrow 2)\right \}.
\label{pg44}
\end{eqnarray}

The renormalization constant for the deuteron state contains  the
nucleon contribution (\ref{pg31}) and the exchange part, $Z_{D} = 2
Z_N + {\tilde Z}_{D}$,
where
\begin{eqnarray}
{\tilde Z}_{D} = - \int \frac{d {\bf p} d {\bf k}}{{(2\pi)}^6}
\Psi_{\cal M}^{+D}({\bf p}) \: \frac{V_\pi({\bf k})}{\omega({\bf k})}
\:
\Psi_{\cal M}^D({\bf p+k}).
\label{pg45}
\end{eqnarray}

We proceed now with an analysis of
the moments of the deuteron SSF. To begin with, let
us define the   isoscalar nucleon SSF,
$g_1^N(x)\equiv (g_1^p(x)+g_1^n(x))/2$.
Then the n-th moment of the  deuteron  solely depends on ${\sf
M}_n(g_1^N)$.
Using
 eqs. (\ref{pg35})-(\ref{pg37}) we get
\begin{eqnarray}
{\sf M}_{n+1}(g_1^N)&=&\bar {\sf M}_{n+1}(g_1^{\bar N}) \left\{1 -
\frac{g^2_\pi}{4m_N^2} \int \frac{d{\bf k}}{{(2\pi)}^3}
\frac{3}{\omega^2({\bf k})} \left ( {\sl N}_n^{(+)}({\bf k}) \: {\bf
k}^2 +
{\sl N}_n^{(-)}({\bf k}) \left( 2k_z^2 - {\bf k}^2 \right) \right)
\right\}
\nonumber \\
&+&\bar {\sf M}_{n+1}(g_1^{\bar N}) \left\{ - Z_N +
\frac{g^2_\pi}{4m_N^2} \int \frac{d{\bf k}}{{(2\pi)}^3}
\frac{3}{2\omega^3({\bf k})} \left( 2k_z^2 - {\bf k}^2 \right) +
n \frac{\delta m}{m_N} \right\}.
\label{pg46}
\end{eqnarray}

The  moments of
the deuteron SSF's are determined by matrix elements similar
to those in eqs. (\ref{pg32})-(\ref{pg34}). In the nuclear case all
the
operators (\ref{pg21})-(\ref{pg23}) contribute to the corresponding
matrix elements.
While evaluating  the matrix
elements one obtains  contributions corresponding to
eqs.~(\ref{pg32})-(\ref{pg34}) computed with the deuteron wave
functions
 (\ref{pg41})  and (\ref{pg42}), the scattering off a bare nucleon
and
 self-energy like corrections (see, also Fig. 2, diagrams (a)). These
diagrams
 provide the contribution
to the Fermi motion of  ``dressed" nucleons.
Besides, there are terms of a pure exchange
origin which reflect the fact that nucleons in the deuteron are
``off-mass-shell", and terms with renormalization and recoil
contributions.

\begin{eqnarray}
&& \hspace*{-.7cm}\frac{1}{2}\left (\frac{M_D}{m_N}
\right )^n\, {\sf M}_{n+1}^D(g_1^D)
= \frac{1}{2} \sum_{{\cal M}=\pm 1}\,
{\cal M} \, \bar {\sf M}_{n+1}^N
\left [
\left (1-\frac{3}{2}P_D\right )
\left (-\frac{\tilde Z_D}{2} +
\frac{{\sf M}_{n+1}(g_1^N) -\bar {\sf M}_{n+1}(g_1^{\bar N})}
{\bar {\sf M}_{n+1} (g_1^{\bar N})}\right )\right .
\nonumber\\[2mm]
&&\hspace*{-.7cm}
+\int \frac{d {\bf p}}{(2\pi)^3} N_n^{(1)}({\bf p},{\bf p})
  \Psi^{*D}_{{\cal M}} ({\bf p})\Sigma^+_{(12)}({\bf p})
\Psi^D_{{\cal M}}  ({\bf p})\nonumber\\[2mm]
 &&-\left .
\int\frac{d{\bf p}d{\bf k}}{(2\pi)^6}
 \Psi^{*D}_{{\cal M}}({\bf p})
 {S_zV_\pi({\bf k})} \left ( \frac{1}{ \omega({\bf k}) } -
 N_n^{(3)}({\bf k}) \right )\Psi^{ D}_{{\cal M}}({\bf p+k})
   \right ],
\label{pg47}
\end{eqnarray}
where
${\bf S}$ is the total spin of the nucleons,
${\bf S} = {1 \over 2} ( {\bfgr\sigma}_1 + {\bfgr \sigma}_2)$,
(with its projection on to the z-axis $S_z$),
$V_\pi({\bf k})$ is the one pion exchange potential,
${\langle S_z \rangle}_{D}  =  \left(1 - {3 / 2}P_D \right)$,
 $P_D$ is the $D$-wave probability
in the deuteron, $\Sigma_{(12)}^+({\bf p}) =
{1\over2} (\Sigma_1^+({\bf p}, {\bf p}) +
\Sigma_2^+({\bf p}, {\bf p}))$
(see eqs.(\ref{pg21}))-(\ref{pg24}))
 and  $N_n^{(3)}({\bf k})$, is defined by
\begin{equation}
N_n^{(3)}({\bf k}) = {1 \over k_z}
\left[
{\left(1 + \frac{k_z}{2m_N} \right)}^n -
{\left(1 - \frac{k_z}{2m_N} \right)}^n
\right].
\label{pg48}
\end{equation}
The expressions in eq. (\ref{pg47}) still contain
the bare moments,
$\bar{\sf M}_n(g_1^{\bar N})$. However they may be expressed trough
the physical
moments by making use of
\begin{eqnarray}
\left ( 1-\frac{3}{2}P_D\right ) \left (
{\sf M}_{n+1}(g_1^N)-\bar{\sf M}_{n+1}(g_1^{\bar N})\right ) =
 \langle S_z\rangle_{D }\left (
{\sf M}_{n+1}(g_1^N)-\bar{\sf M}_{n+1}(g_1^{\bar N})\right )
&&\nonumber\\[1mm]
\approx
 \langle \Sigma_{(12)}^+\cdot N_n^{(1)}({\bf p})\rangle_{D}
\left ({\sf M}_{n+1}(g_1^N)-\bar{\sf M}_{n+1}(g_1^{\bar N})
\right ).&&
\label{pg49}
\end{eqnarray}
In obtaining expression (\ref{pg49}), we have used that fact that
$\left ({\sf M}_n(g_1^N)-\bar{\sf M}_n(g_1^{\bar N})\right )
\sim g_\pi^2$;
then the moments of the  deuteron, up to $g^2_\pi$ terms,
contain only well-defined quantities, viz
\begin{eqnarray}
{1 \over 2}
{\left(\frac{M_D}{m_N}\right)}^{n}&\times&{\sf M}_{n+1}(g_1^D)=
{\sf M}_{n+1}(g_1^N) \: \Delta Z
\label{pg50} \\
&+& {\sf M}_{n+1}(g_1^N)
\int \frac{d{\bf p}}{{(2\pi)}^3} \:
  {\sl f}_D^{IA}({\bf p}) \:
{\left(1 + \frac{p_z}{m_N} + \frac{{\bf p}^2}{2m_N^2} \right)}^n
 \nonumber \\
&+&
{\sf M}_{n+1}(g_1^N)
\int \frac{d{\bf p} d{\bf k} }{{(2\pi)}^6} \:
  {\sl f}_D^{int}({\bf p, k}) \:
{1 \over k_z}
\left[
{\left(1 + \frac{k_z}{2m_N} \right)}^n -
{\left(1 - \frac{k_z}{2m_N} \right)}^n
\right], \nonumber
\end{eqnarray}
where
\begin{eqnarray}
{\sl f}_D^{IA}({\bf p})&=&{1\over 2} \sum_{{\cal M}=\pm 1}\,
  {\cal M} \, \Psi_{{\cal M}}^{*D}({\bf p})
\left(
\frac{{\bf S} \cdot {\bf p}}{m_N} + S_z +
\frac{{\bf S} \cdot {\bf p}}{2m_N^2}\: p_z \right)
\Psi_{{\cal M}}^D({\bf p}),
\label{pg52}\\[1mm]
{\sl f}_D^{int}({\bf p,k})&=&{1\over 4} \sum_{{\cal M}=\pm 1}\,
 {\cal M} \,\Psi_{{\cal M}}^{*D}({\bf p}) \:
  \left\{ S_z ,\: V_\pi({\bf k}) \right\}   \:
\Psi_{{\cal M}}^D({\bf p+k}),
\label{pg53}
\end{eqnarray}
where $\left\{S_z, V_\pi \right\}$ stands
for  the anticommutator of $S_z$ and  $V_\pi$ and
$\Delta Z$, which is the difference between the renormalization
 and recoil contributions, has the same origin as in the case
of nucleons and reads as follows
\begin{eqnarray}
 2\Delta Z = {1\over 2}\sum_{{\cal M}=\pm 1}\,
 {\cal M} \int \frac{d{\bf p} d{\bf k} }{{(2\pi)}^6}
\Psi_{{\cal M}}^{*D}({\bf p})
\left\{
\frac{ {\langle S_z \rangle}  - S_z}{2\omega({\bf k})},
 V_\pi({\bf k})
\right\}
\Psi_{{\cal M}}^D({\bf p + k}).
\nonumber\\
\label{pg54}
\end{eqnarray}

 Figure 2 illustrates the dressing  of the moments of the
 deuteron. The sum of diagrams (a) gives the impulse
approximation, that is, the scattering off a physical, already
dressed nucleon. Comparing these diagrams with those
in Fig. 1, it is seen that in the impulse approximation the
helicity amplitude on the  deuteron
is determined by the scattering amplitudes
off free physical nucleons, which
ought to be fixed from other experiments. In
this context, the present approach allows one to avoid
the problem as to what amplitudes should be associated with
``off-mass-shell'' nucleon in impulse approximation
(see, for instance, Ref.\cite{forest}).
In our approach the binding effects are taken into account by
terms which are of a pure exchange origin (Fig. 2, diagrams (c))
which contain implicitly, via the potential $V_\pi({\bf k})$,
 the contribution of the meson degrees of freedom. The diagrams
(b) are the recoil contribution terms.

Applying the inverse Mellin transform to (\ref{pg50})
and omitting the $\Delta Z$ part, the deuteron
structure function $g_1^D$ can be obtained
in the convolution form
\begin{eqnarray}
{1 \over 2} g_1^D(x) = \int \limits_{x}^{M_D/m}
\frac{dy}{y} \: g_1^N \left( \frac{x}{y} \right)
\left(   {\sf f}_D^{IA}(y)
+   {\sf f}_D^{int.}(y) \right),
\label{pg55}
\end{eqnarray}
where the distribution functions
$  {\sf f}_D^{IA}(y)$ and $  {\sf f}_D^{int.}(y)$ are given by
\begin{eqnarray}
  {\sf f}_D^{IA}(y)&=&
\!\!
\int \frac{d{\bf p}}{{(2\pi)}^3} \:
  {\sl f}_D^{IA}({\bf p}) \: \delta \left( y - 1 - \frac{p_z}{m_N}
- \frac{{\bf p}^2}{2m_N^2} \right), \label{pg56} \\
  {\sf f}_D^{int.}(y)&=&
\!\!
\int \frac{d{\bf p} d{\bf k} }{{(2\pi)}^6}
  {\sl f}_D^{int}({\bf p, k}) {1 \over k_z}
\left[
 \delta \left(1 - y + \frac{k_z}{2m_N} \right)  -
 \delta \left(1 - y - \frac{k_z}{2m_N} \right)
\right] \theta(y),
\label{pg57}
\end{eqnarray}
where ${\sf f}^{IA}$ corresponds to the impulse approximation, or
Fermi motion
correction, with ``on-mass-shell" nucleons, and ${\sf f}^{int}$
accounts for the binding of the nucleon inside the deuteron.

Equations (\ref{pg50}) and (\ref{pg55})-(\ref{pg57}) are
the basic result for the determination of the moments and
the SSF's of the deuteron within the OPE-OBE
approach.
It will be shown later on that eqs. (\ref{pg55})-(\ref{pg57})
lead, if proper assumptions are made, to the
phenomenological convolution model approach used
in refs.\cite{ciofi2}.
It is worth recalling here the problem as to
whether
the so-called flux factor has to be considered in the
convolution formula for polarized deep inelastic
scattering \cite{ciofi3,francfurt}.
In our approach the non-relativistic flux factor
$\sim (1+p_z/m_N)$ comes automatically, as it
 can be seen from eq. (\ref{pg52}).

Formulae (\ref{pg50}) - (\ref{pg57})
 have been obtained for the pseudoscalar isovector coupling.
In this case the deuteron wave function
$\Psi^D({\bf p})$  appears to be the solution of
the Schr\"odinger equation with the one pion exchange
  $NN$-potential $V_\pi({\bf k})$.
Obviously, this wave function and the one pion exchange potential
are not yet sufficient to describe the properties of the  deuteron.
To this end it is necessary  to take into
 account other mesons contributing to the OBE potential,  viz.
the $\sigma,\, \omega,\, \rho, \, \eta$ and $\delta $
 mesons~\cite{brown}. Including these mesons in our
approach leads to contributions
similar to (\ref{pg50}), (\ref{pg57}),
except that the wave function
$\Psi^D({\bf p})$ is replaced by
the solution of the Schr\"odinger equation
with the full OBE potential.
The convolution formula can be written in a more
compact form by
expanding the $\delta$ functions in
(\ref{pg57}) and retaining terms up to $k^2_z/m_N^2$; one gets
\begin{eqnarray}
{1 \over 2} g_1^D(x) = g_1^{IA}(x) -
\frac{d}{dx} \left (
x g_1^N(x) \right )\cdot \frac{
 {\langle  S_z, V_{OBE}({\bf k})  \rangle}_D }{ m_N}
\label{pg58}
\end{eqnarray}
where
\begin{eqnarray}
g_1^{IA}(x)=\int\limits_x^{M_D / M} {dy \over y}g_1^N \left
({x \over y} \right){\sf f}^{IA}(y)~~,
\label{pg 58bis}
\end{eqnarray}
and  $\langle S_z V_{OBE} \rangle_{D}$ is the spin-weighted
mean value of the
potential of the  nucleon   in the polarized
deuteron.

The second term in eq. (\ref{pg58}) is the correction to
the impulse approximation due to the binding of
nucleons. It can be seen, that this contribution is
small ($\sim \, \langle\, V_{OBE} \,\rangle/m_N$)
and  depends on the behavior of the nucleon
structure function $g_1^N(x)$ and its first derivative.

\subsection{The $^3He$}
In this section the Compton helicity amplitude
$h_{+ \: {\cal M},+ \: {\cal M}}$ pertaining to the
scattering of the virtual helicity $(+)$ photon off the polarized
(with projection ${\cal M} = \pm \frac{1}{2}$) $^3He$ nucleus
will be presented.
 Using the Tamm-Dancoff decomposition for the state vector
$| ^3 He \rangle_{{\bf Q}=0, {\cal M}}$
and the operators $O^\lambda_t(n)$,
the following form
for the helicity amplitude can be obtained
\begin{eqnarray}
&&\hspace*{-7mm}
h_{+ \: {\cal M},+ \: {\cal M}}=
 - \frac{2M_{^3He}}{m_N} \sum_{ n=0,2,\ldots}^{\infty}
 {\left( \frac{1}{x}\right)}^{n+1}
 \int  dV_{^3He}\Psi_{\cal M}^{^3He*}({\bf p}_1, {\bf p}_2,
  {\bf p}_3) \left \{
  \left [ \int \frac{d{\bf k} }{(2\pi)^3}
\frac{g^2_\pi}{4m_N^2} \frac{1}{\omega^2}
  \sum_{i=1}^3
  \phantom{ \int \frac{d{\bf p}_1}{(2)^6} }\right .\right .
  \nonumber\\[1mm]
&&\hspace*{-10mm}
\left \{
   3 {\bf k}^2 \sigma_z^{(i)} \hat C_n^{(i)}
\left ( \frac{1}{2 \omega} -  {\sl N}^{(+)}_n({\bf k})\right )
+
\left (2 k_z \,(\bfgr \sigma^{(i)}\,{\bf k}) -
{\bf k}^2\sigma_z^{(i)}\right )\: {\bfgr \tau^{(i)}}
\hat C_n^{(i)}{\bfgr \tau^{(i)}}
\left (  \frac{1}{ 2\omega} - {\sl N}^{(-)}_n({\bf k})\right )
\right \}\label{pg59}
\\&&\hspace*{-10mm}
\left .   +  \left( 1 - Z_{^3He} \right)\sum_{i=1}^3
 {\sl N}_n^{(1)}({\bf p}_i) \Sigma_i^+ ({\bf p}_i)
 \hat C_n^{(i)} \: \right ]
 \Psi_{{\cal M}}^{^3He}({\bf p}_1, {\bf p}_2, {\bf p}_3)
\label{pg60}\\[1mm]
&&\hspace*{-10mm}
\left.
+\int \frac{d{\bf k} }{(2\pi)^3}
 \: \sum_{i\: < \:j}
\left[
\left\{ {V_{\pi i j}}, S_{z ij} \right\} \left(
\frac{1}{\omega} +{\sl N}^{(3)}_n({\bf k})\right )+
  \frac{V_{\pi i j}}{\omega} \sigma_z^{(k)} \hat C_n^{(k)}\right] \:
\Psi^{He}_{{\cal M}}
 ({\bf p}_i + {\bf k},{\bf p}_k,{\bf p}_j - {\bf k})\right \}
\label{pg61}
\end{eqnarray}
where
$
 S_{z ij} = {1\over 2}
\left( \sigma_z^{(i)} \hat C_n^{(i)} + \sigma_z^{(j)}
 \hat C_n^{(j)}\right)$,
  $ dV_{^3He}\equiv \prod\limits_{i=1}^3
  \displaystyle\frac{d{\bf p}_i}{(2\pi)^6}
\delta ( {\bf p_1}+ {\bf p_2}+ {\bf p_3})$
and in eq. (\ref{pg61})
$i,\, j$ list two nucleon pairs exchanging a meson while the
remaining
 $k$-th nucleon interacts with the incoming lepton,
$k \neq i, j$.
In eq. (\ref{pg59})  the reader may easily recognize
the corresponding ``dressing'' part for bare nucleons,
eq. (\ref{pg36}) and (\ref{pg37}), which being, included into
eq. (\ref{pg60}), gives the impulse approximation with the
{\em physical} nucleons. The last term in eq. (\ref{pg61})
 and the  term depending upon
the renormalization constant $Z_{^3He}$
in eq.  (\ref{pg60}) correspond to the recoil and
renormalization
contributions, similar to those obtained in the deuteron
case.
The moments of the $^3He$ SSF's can easily be obtained
by
recalling eqs. (\ref{pg15}) and (\ref{pg38}): the inverse
Mellin transform gives the SSF in the convolution form
\begin{eqnarray}
g_1^{^3He}(x) =
\int \limits_{x}^{M_{^3He}/m_N}
\frac{dy}{y} \: \left[
g_1^p \left( \frac{x}{y} \right) {\sf f}^p_{^3He}(y) +
g_1^n \left( \frac{x}{y} \right) {\sf f}^n_{^3He}(y)
\right],
\label{pg3t}
\end{eqnarray}
where the effective distribution functions ${\sf f}^{p,n}_{^3He}(y)$
of nucleons
in $^3He$ contain contribution from the
Fermi motion of ``on-mass-shell'' nucleons (impulse approximation)
and from the nuclear binding
\begin{equation}
{\sf f}^{p,n}_{^3He}(y) = {\sf f}_{IA}^{p,n}(y) + {\sf
f}_{int.}^{p,n}(y),
\end{equation}
where
\begin{eqnarray}
&&{\sf f}_{IA}^{p,n}(y)=
\int dV_{^3He} \:
n_{\|}^{p,n} ({\bf p}_1,{\bf p}_2,{\bf p}_3)
\: \delta \left( y - 1 - \frac{p_{{z_1}}}{m_N}
- {{\bf p}_1^2 \over 2m_N^2} \right), \label{pg3ia} \\
&&
{\sf f}_{int.}^{p,n}(y)=\nonumber\\
&&
\int \frac{dV_{^3He}d{\bf k}}{{(2\pi)}^3}
  n_{int.}^{p,n}({\bf p}_1,{\bf p}_2,{\bf p}_3, {\bf k})
{ \theta(y)  \over k_z}
\left[
 \delta \left(1 - y + \frac{k_z}{2m_N} \right)  -
 \delta \left(1 - y - \frac{k_z}{2m_N} \right)
\right] \nonumber \\ \label{pg3int}
\end{eqnarray}
with the spin-dependent momentum distributions
$n_{\|}^{p,n}$ and  $n_{int.}^{p,n}$  being defined by
\begin{eqnarray}
&&n_{\|}^{p,n} ({\bf p}_1,{\bf p}_2,{\bf p}_3) =
 \sum_{{\cal M}=\pm  1/2 ,i}\,
 {\cal M} \, \Psi_{\cal M}^{^3He*}
\Sigma^+_{i}({\bf p}_i)\left [ (1\pm \tau_3(i)) / 2 \right ]
\Psi_{{\cal M}}^{^3He},
\label{pg3fia}\\[2mm]
&&n_{int.}^{p,n}({\bf p}_1,{\bf p}_2,{\bf p}_3, {\bf k}) =
\label{pgrfint}\\
&&=
 \sum_{{\cal M}=\pm  1/2, i\neq j}\,
 {\cal M} \, \Psi_{{\cal M}}^{^3He*}( {\bf p}_1,{\bf p}_2,{\bf p}_3 )
\:
  V_{\pi, ij} S_{z ij}
\left [ (1\pm \tau_3(j))/ 2 \right ]
\Psi_{{\cal M}}^{^3He} ({\bf p}_i + {\bf k},{\bf p}_k,{\bf p}_j -
{\bf k}).
\nonumber
\end{eqnarray}
 Eq.~(\ref{pg3t}) can be cast in a more
manageable form by expanding
the $\delta$ functions as in the case of the deuteron
\begin{eqnarray}
  &&g_1^{^3He}(x) =
\int \limits_{x}^{M_{^3He}/m_N}
\frac{dy}{y} \left[
 g_{1}^{p} \left (  \frac{x}{y}\right ) {\sf f}_{IA}^p(y)+
 g_{1}^{n} \left (  \frac{x}{y}\right ) {\sf f}_{IA}^n(y)
\right]
\label{pg62}\\[1mm]
&&
 -\frac{d}{dx}\left ( xg_{1}^{p}(x)\right )
 \frac{2\langle\sigma_z^p\cdot V_{OBE}^p\rangle_{^3He}}{m_N}
 \,-
 \frac{d}{dx}
\left ( xg_{1}^{n}(x)\right )
\frac{\langle\sigma_z^n\cdot  V_{OBE}^n\rangle_{^3He}}{m_N},
\nonumber
\end{eqnarray}
where $\langle\sigma_z^i\cdot  V_{OBE}^i\rangle_{^3He}$ is the
spin-weighted
mean value of the
potential of the  nucleon ( $i=p,n$) in the polarized
$^3He$.

\subsection{Generalization to heavy nuclei}
Let us generalize our approach to heavy nuclei. To this end, let us
first
discuss the unpolarized case. The comparison will be carried up to
a given order in $(p/m_N)^n$, namely to $n=2$.

Guided by the results for $A=2$, one can write for a generic
isoscalar
nucleus $A$~\cite{kaz}
\footnote{Note that in eq. (16)
of Ref.~\cite{kaz}
the ``plus" sign in Eq. (16) should be
replaced by a ``minus" sign.}
\begin{equation}
F_2^A(x)=F_2^{IA}(x)-x{ dF_2^N(x) \over dx}
\frac{ \langle V \rangle_A}{m_N}~~,
\label{p1}
\end{equation}
where
\begin{equation}
F_2^{IA}(x)=\int F_2^N \left( {x \over y} \right) f_A^{IA}(y) dy,
\label{c21}
\end{equation}
and
\begin{equation}
f_A^{IA}(y)=\int \left( 1 + {p_z \over m_N} \right) n(p)
\delta \left( y - \left[ 1+{{\bf p}^2 \over 2m_N} + {p_z \over m_N}
\right]
\right) d{\bf p}.
\label{c31}
\end{equation}
In Eq. (\ref{p1}) $\langle V \rangle_A$ is the mean potential energy
of the nucleon interacting with the lepton; it is linked to the
nuclear
 mean potential energy per nucleon
$\langle V \rangle \equiv \langle \Psi_A| \sum_{i<j}v_{ij}|
\Psi_A \rangle /A$ by the relation
$\langle V \rangle_A=2 \langle V \rangle$. The well known
relationships between
the total energy per nucleon
$\epsilon_A=E_A/A$,
the mean nucleon kinetic energy
$\langle T \rangle=\langle \Psi_A|
\sum_{i}t_{i}|
\Psi_A \rangle /A$ and
the mean removal energy per nucleon $\langle E \rangle$
are given by
$\epsilon_A=\langle T \rangle + \langle V \rangle$
and $\langle E \rangle=
 2 |\epsilon_A| +  (A-2) \langle T \rangle/(A-1)$.

Now we turn to the deuteron eqs.~(\ref{pg52}), (\ref{pg53}) and
(\ref{pg55})--(\ref{pg57}). In the definition of the distribution
functions
 in impulse approximation, eq.~(\ref{pg52}), it is easy to identify
 the non-relativistic analogue of the nucleon spin vector
 $ S_z + ({\bf S\cdot p})/2m_N^2$, which may
be obtained by applying the Lorentz boost operator
 to the spin vector of a nucleon at rest. In the non-relativistic
limit the spin vector
is the same in
any reference frame. Consequently, in the extreme
non-relativistic limit it should be replaced by $S_z$ only.
Furthermore, in the matrix element
of $({\bf S\cdot p})/m_N$ in eq.~(\ref{pg52}) only the
$z$-components of the scalar product $ ({\bf S\cdot p})$ give
contribution as it can be checked by direct computation.
So that in this case the distribution function
 $f^{IA}({\bf p})$ becomes
\begin{eqnarray}
{\sl f}^{IA}({\bf p})&=&{1\over 2} \sum_{{\cal M}=\pm 1}\,
  {\cal M} \, \Psi_{{\cal M}}^{*D}({\bf p})
\left(1+
\frac{p_z}{m_N}  \right)\, S_z\, \Psi_{{\cal M}}^D({\bf p}).
\label{gen1}
\end{eqnarray}
Expanding the $\delta$-function in (\ref{pg57}) around the
``on-mass-shell"
$y$,  $\delta (y-1-{\bf p}^2/2m_N^2- p_z/m_N)\approx
\delta(1-y)-\delta'
(1-y)({\bf p}^2/2m_N^2+ p_z/m_N)$  and substituting in eq.
(\ref{pg57})
the difference of two $\delta$-functions by its first derivative we
get
\begin{eqnarray}
&&f^{N/D}_\|(y)\equiv
f^{IA}(y) + f^{int.}(y) \approx  {1\over 2} \sum_{{\cal M}=\pm 1}
\,  {\cal M} \,\int\frac{d{\bf p}}{(2\pi)^3}
\Psi^{*D}_{\cal M}({\bf p}) \left ( 1+\frac{p_z}{m_N}\right )S_z
\Psi^{D}_{\cal M}({\bf p})\nonumber \\[2mm]
&& \times \left [
\delta (1-y)-\delta'(1-y)({\bf p}^2/2m_N^2+ p_z/m_N+\varepsilon_D/m_N
-
\langle T \rangle /m_N ) \right ].
\label{gen2}
\end{eqnarray}
In deriving eq.~(\ref{gen2}) we used the Schr\"odinger equation to
express $\langle   S_z, V  \rangle$ through the deuteron binding
energy $\varepsilon_D$ and the kinetic energy $\langle T \rangle$ of
nucleons.
It can be easily shown that eq.~(\ref{gen2}) can be cast in the form
\begin{eqnarray}
f^{N/D}_\|(y) = \int\frac{d{\bf p}}{(2\pi)^3}
dE \,{\cal P}^D_{\|}({\bf p},E)\left (
1+\frac{p_z}{m_N}\right )
\delta\left( y-\left [{m_N-E-{\bf p}^2/2m_N + p_z \over m_N} \right ]
\right ),\label{gen3}
\end{eqnarray}
where we have introduced the deuteron spin dependent spectral
function
\begin{eqnarray}
{\cal P}^D_{\|}({\bf p},E)
\equiv\Psi^{*D}_{{\cal M}=1}({\bf p})\,
S_z\,\Psi^{D}_{{\cal M}=1}({\bf p})\,
\delta(E-|\varepsilon_D| )
\label{gen4}
\end{eqnarray}
giving the probability to find in the deuteron a nucleon with
momentum
${\bf p}$, removal energy $E=|\varepsilon_D|$ and spin projection
$S_z$.
In order to generalize eq. (\ref{gen3}) to a heavy nucleus, we notice
that the quantity $m_N-E-{\bf p}^2/2m_N$ appearing in the $\delta$
function of eq. (\ref{gen3}) is nothing but the time component
of the four momentum of an off--shell nucleon in the deuteron, $
viz.$
$p_0=M_D-\sqrt{{\bf p^2}+m_N^2}\simeq m_N -|\varepsilon_D| -
{\bf p}^2/2m_N$.

Formula (\ref{gen3}) can therefore be generalized to the case of
any nuclear mass number $A>2$ by substituting the ``deuteron
spin--dependent spectral function"
${\cal P}^D_{\cal M}({\bf p},E)$ with the corresponding
nuclear spin dependent spectral function~\cite{ciofi1,donelly}.
Then eq.~(\ref{gen3}) exactly coincides with the phenomenological
convolution approach used in ~\cite{ciofi1,ciofi2}
(see next Section).

\narrowtext
\subsection{Comparison with the conventional convolution approach}
Using the concept of spin--dependent spectral function ~\cite{ciofi1},
the SSF of $^3He$ has been recently calculated ~\cite{ciofi2} within
the so called convolution approach
in which the lepton is assumed to interact with off shell nucleons
with four momentum $p\equiv(p_0,{\bf p})$, with $p_0=M_A-\sqrt{
(E-m_N+M_A)^2+{\bf p}^2}$, where $M_A$ is the sum of the target
nucleus and $E=M_{A-1}+m_N-M_A+E_{A-1}^*$ is the nucleon removal
energy with $M_{A-1}$ being the mass of the spectator $A-1$ system
and $E_{A-1}^*$ its intrinsic excitation energy.
Since in our calculations we have used the
results of Ref.~\cite{ciofi1}
it is worth comparing
the two approaches from a formal point of view.
To this end, we start with the unpolarized case.
Then we can use the analogous of (\ref{pg58}) for an isoscalar
nucleus in the unpolarized case and take its generalization to any
value
of $A$, as we did in the previous section obtaining
eqs. (\ref{p1}) -- (\ref{c31}).
 The convolution formula
for off shell nucleons reads as follows ~\cite{ciofi3}:
\begin{equation}
F_2^A(x)=\int F_2^A \left({x \over y} \right) f_A (y) dy,
\label{c4}
\end{equation}
where
\begin{equation}
f_A(y)=\int \left( {p^+ \over m_N } \right) P(|{\bf p}|,E)
\delta \left( y - {p^+ \over m_N} \right)
d{\bf p} dE~~,
\label{c5}
\end{equation}
with $p^+=p_0+p_z$ and $P(|{\bf p}|,E)$ being the unpolarized
nucleon spectral function. It can be seen that whereas in eq.
(\ref{pg58}) the binding effect is
explicitly displayed through the mean potential energy
$ \langle V \rangle $, in the conventional convolution approach
the binding effects
are hidden in the definition of the light
cone momentum distribution $f_A(y)$. Let us however write down the
latter
in the order $p^2/m_N^2$ as in Eq. (\ref{pg58}). To this end, we
expand the
$\delta$ function in Eq. (\ref{c5}) around the point
$y - \left[ 1+\displaystyle\frac{p^2}{ 2m_N} +
\displaystyle\frac{p_z}{m_N} \right]$ and keep only
terms of the order ${\bf p}^2/m_N^2$ ; we get
\begin{eqnarray}
f_A(y) &\simeq& \int \left( 1 + {p_z \over m_N} \right) n(|{\bf p}|)
\nonumber \\
&\times&
\delta \left( y - \left[ 1+{p^2 \over 2m_N} + {p_z \over m_N} \right]
\right) d{\bf p}
\nonumber \\
&+&
 {\langle E \rangle +  \langle T \rangle
\,A/(A-1)\over m_N}
\delta'(y-1)
\label{c6}
\end{eqnarray}
where
$ n(|{\bf p}|)=\int P(|{\bf p}|,E) dE $ and
$ \langle E \rangle= \int E P(|{\bf p}|,E) d {\bf p} dE$.
Placing (\ref{c6}) in (\ref{c4}) and using the relationship
$ \langle E \rangle + \langle T \rangle = 2  \langle V \rangle $
we recover Eq. (\ref{p1}). Thus we have demonstrated the
convolution formulae arising from the OPE and the conventional
treatment of unpolarized DIS coincide up to the order $ p^2/m_N^2$;
likewise we have shown that at this order the binding effect in
the latter approach arises from the average potential energy of the
nucleon
hit by the incoming lepton. Let us turn now to the polarized
case and let us analyze the $^3He$ case. The conventional approach
yields ~\cite{ciofi2}
\begin{eqnarray}
g_1^{^3He}(x) = \sum_{i=n,p} \int_x^A {dy\over y}
g_1^i \left( {x \over y}
\right) G^i(y)~~, \label{pgn1}
\end{eqnarray}
where the light cone momentum distribution $G^i(y)$
is given by the following expression:
\begin{eqnarray}
G^i(y)= \int dE \int d{\bf p} P_{\|}^i({\bf p},E)
\delta \left( y-{{p_0+p_z} \over m_N} \right)~,
\label{pgn2}
\end{eqnarray}
where $E=M_D+m_N-M_{^3He}+E_{D^*}$ is the nucleon removal energy
($E_{D^*}$ being the energy of the spectator np pair in the
continuum),
$p_0=M_{^3He}-[(E-m_N+M_{^3He})^2+|{\bf p}|^2]^{1\over 2}$ is the
energy of a bound ``off-mass-shell" nucleon, and
$P_{\|}^i({\bf p},E)$ is
the spin dependent spectral function
(cf. eqs. (9) and (16) of Ref. \cite{ciofi2}).
The integral of
the spin dependent spectral function represents the spin dependent
momentum distribution
\begin{eqnarray}
n_{\|}^i({\bf p})=\int dE P_{\|}^i({\bf p},E)~~.
\label{pgn3}
\end{eqnarray}
 By expanding the $\delta$ function in (\ref{pgn3}) around
$ y - \left[ 1+\displaystyle\frac{p^2}{ 2m_N} +
\displaystyle\frac{p_z}{ m_N} \right]$ and considering the
recoil of the  two body system non relativistically
($E_R= p^2/4m_N$), and taking into account the differences between
the proton and neutron spectral functions (see \cite{ciofi3}),
it can be shown that (\ref{pg3t}) and  (\ref{pg62}) are recovered.

\section{Results of calculations}
\setcounter{equation}{0}
\subsection{The Deuteron}
 For explicit numerical calculation of the nuclear SSF,
  $g_1^D(x)$ and $g_1^{^3He}(x)$,  one needs, as it can
   be seen from eqs. (\ref{pg55}) and
 (\ref{pg62}), a suitable  parameterization of the isoscalar
  nucleon SSF $g_1^N(x)$.
Within the present  approach
 this function contains
 all the information about the Wilson's coefficients and the
  influence of the meson cloud on bare nucleon; moreover,
according to the main assumptions
of the effective  meson-nucleon theory,
it includes all the dynamic at distances shorter than
the core  of OBE-potential.
The parameterization of $g_1^N$ should be fixed from the experiment;
in principle, there exist nowadays experimental data
of $g_1$ both
on the
 proton~\cite{spin} and the neutron~\cite{slac}
 structure functions,
 but they are not yet fully complete, especially
at very small values of $x$. In this interval some assumptions
 about the behavior of the
 nucleon  structure function are
 unavoidable. Moreover, the choice
  of the isoscalar structure function $g_1^N(x)$
  determines whether  the Bjorken and Ellis-Jaffe sum
 rules will be fulfilled or not.
 In this sense our results depend on the parameterization of the
 nucleon SSF, however this dependence is found to be
  insignificant (see below).
 We have chosen here two different parameterizations
 of the nucleon SSF, describing
 quite well the  EMC data on
 proton and satisfying  the Bjorken sum rule.
 We use the parameterization
 from Ref.~\cite{shaef} as a basic input in all our calculations,
and in order
to analyse the dependence of the results from the chosen
parameterization
we  use the parameterization from Ref.~\cite{dor}.

In our numerical calculations for the deuteron we use
 the wave function obtained from the Bonn interaction,
yielding $P_D
 \approx 0.0428$~\cite{bonn}.
 Figure 3  displays the numerical estimate
 of the ratio $R_D=g_1^D/g_1^N$, which illustrates the effects of
nuclear
structure on $g_1^D$.
The results presented in this figure, deserve the following comments:
i) within the impulse approximation (dotted line; eq. (\ref{pg55})
without ${\bf f}^{int}$), the ratio in the interval
$0.2 < \, x\, <\,0.7$
is governed by the
destructive contributions of the $D$-wave admixture
which generates a polarization of the deuteron along
the  $z$-direction even if the nucleons have their spins
aligned in the direction opposite to the polarization.
Thus it can be concluded that
the effect of the $D$ wave
is the most relevant nuclear contribution within the impulse
approximation;
ii) as for
the interaction term,
according to
(\ref{pg58}), the binding
 corrections to the deuteron SSF is
governed by
$\langle  S_z,V_{OBE} \rangle_D$.
Let us first of all estimate the contribution due to pions; a direct
numerical computation gives
\begin{eqnarray}
 \langle S_z,V_{\pi}\rangle_D \sim - 5 MeV
 \label{estim1}
\end{eqnarray}
(if model ambiguities given e.g. by different choices for the
form-factors and
 wave functions are considered, one gets
 $\langle  S_z,V_{\pi} \rangle_D \sim - (3\div 5) \, MeV$).
It can be seen from Fig. 3
that the effects from binding due to $\pi$ mesons
is similar to the one occurring in the unpolarized case,
but significantly smaller.
It can also be seen
that the results are not sensitive to the
parameterization  of the nucleon SSF
(cf. long and short dashed waves).
We turn now to the estimate of the contributions due to other mesons,
although it could be
anticipated that
the pions give the dominant contribution to the
binding effects, since the pion contribution
in the deuteron is the
most significant one ~\cite{kunucl,mel}.
In order
to estimate the most general
boson contribution
to the deuteron SSF
one should, in principle, calculate the matrix elements
$\langle  S_z,V_{OBE} \rangle_D $ for all kinds of bosons
considered by the Bonn potential model.
To this end one can use the Schr\"odinger equation to get
  \begin{equation}
\langle  S_z, V_{OBE}  \rangle = \left ( 1-\frac{3}{2}P_D \right )
 \varepsilon_D -
\left (\langle T \rangle_{0} - {1 \over 2}
\langle T \rangle_{2} \right ),
\label{pg64}
\end{equation}
where  $\langle T \rangle _{0,2}$
are the mean values of the nucleon kinetic energy in the
$S-$ and $D-$ waves (for the deuteron wave function
of the Bonn  potential we have
$\langle T \rangle_{0}\approx 10.2$ MeV
and $\langle T \rangle_{2} \approx 4.4$ MeV). Taking into account
Fermi-motion and the full interaction effects
by eq. (\ref{pg64})
results in a
EMC--like effect, as in the unpolarized case (full curve in Fig. 3).
By comparing the dotted curve (only Fermi motion) with the full one
(Fermi
motion plus binding), it can be concluded that
the main nuclear structure effect in
the deuteron  SSF comes from the
presence of the $D$-wave,
with the binding effect remaining a rather small correction.
 Note that all curves in the Fig.~3 are not shown at values of $x$
smaller then $x \approx 0.2$. The
reason is that realistic parameterizations of the nucleon SSF have
nodes at small $x$, which lead to ``poles''
in the ratio $g_1^D/g_1^N$ simulating nuclear effects.
At values of $x$ smaller then these poles,
all curves in Fig.~3 tend to the
limit $\left ( 1-\frac{3}{2}P_D \right )$ as $x \to 0$ (this is
 a general result for all kinds of
parameterization no more singular at the
 origin than $1/x$).

Figures 4a and 4b  display  the calculations of the absolute
 values of the deuteron structure function and the comparison
 with the SMC experimental data:
a good agreement between our results and the
experimental data can be observed.
 The numerical estimate of the first moment of $g_1^D(x)$ within
 our approach, $\int dx g_1^D(x) \approx 0.03 $, is also in an
 agreement with the experimental result
 $\int dx g_1^{D(SMC)}(x) = 0.023\pm 0.02 \pm 0.015$~\cite{smc}.

\subsection{$^3He$}
As is well known ~\cite{fri} the interest in polarized
$^3He$ targets stems from the fact that such a system can be
considered to a large extent as an effective polarized neutron
target.
As a matter of fact,
the two protons in  $^3He$ are mostly in the singlet $^1S_0$
configuration so that the polarization of the $^3He$ is
mainly determined by the polarization of the neutron~\cite{vol,fri}.
For such a reason DIS experiments off polarized $^3He$ are aimed at
obtaining information on the neutron SSF's.
However, the non vanishing proton contribution to the total
polarization
and nuclear structure effects
could in principle hinder the direct extraction of the neutron SSF.
We have calculated the $^3He$ spin structure function $g_1^{^3He}$
(eq. (\ref{pg3t})) using the same approximation as in the case of
$^2$H, i.e. by using eq. (\ref{pg62}).

The interaction term, obtained using the Schr\"odinger equation and
a three--body wave function containing $S,S'$ and $D$ waves, reads as
follows
\begin{eqnarray}
 \langle\sigma_z^p\cdot V_{OBE}^p\rangle_{^3He}
&=& 2\left (\frac{2}{3}P'_S-\frac{2}{3}P_D\right )
\cdot\overline{\varepsilon}_{^3He}
\nonumber \\
&-&
2\left (\frac{2}{3}\langle T\rangle_{S'}-\frac{2}{3}\langle
T\rangle_{D}\right)
\nonumber\\
&=&2p^p\cdot\overline{\varepsilon}_{^3He}-2\langle
T\rangle_{\|}^p
 \label{uno}\\
 \langle\sigma_z^n \cdot V_{OBE}^n\rangle_{^3He}
&=& 2\left (P_S+\frac{1}{3}P_{S'}-P_D \right )
\cdot\overline{\varepsilon}_{^3He}
\nonumber \\
&-&
2\left ( \langle T\rangle_{S}+\frac{1}{3}\langle T\rangle_{S'}-
\langle T\rangle_{D}\right )
\nonumber\\
&=&
2p^n\cdot\overline{\varepsilon}_{^3He}-2\langle T\rangle_{\|}^n
\label{nuo1}
\end{eqnarray}
where
$P_{\|}^i=\int n^i_{\|}({\bf p})d{\bf p}/(2\pi)^3$,
$\langle T\rangle^i_\|=\int n^i_{\|}({\bf p}){\bf p}^2/2m_N\,d{\bf
p}/(2\pi)^3$
and $\overline{\varepsilon}_{^3He}$ is the mean value of the
binding  energy per nucleon in $^3He$.

The spin-dependent momentum distributions, the effective nuclear
polarizations and the mean values of the kinetic energies have been
taken
from Ref.~\cite{ciofi2}, where these quantities have been calculated
from
the spin-dependent spectral function obtained from the Reid Soft Core
interaction. The results are
\begin{equation}
\langle\sigma_z^p\cdot V_{OBE}^p\rangle_{^3He}\approx 2.4 MeV,
\label{unoz}
\end{equation}
\begin{equation}
\langle\sigma_z^n\cdot V_{OBE}^n\rangle_{^3He}\approx -17.8 MeV.
\label{due}
\end{equation}

In Fig. 5 the ratio
$R_{^3He}={ g_1^{^3He}/ g_1^n}$
calculated using the proton and neutron spin structure functions from
Ref.
{}~\cite{shaef} and ~\cite{dor} is presented,
whereas in Fig. 6 the SSF $g_1^{^3He}$ is compared with the free
neutron
SSF.
 It can be seen that,
as in the deuteron case, the contribution
of binding effects is rather small.
It should be pointed out that
the results presented in Figs. 5 and 6 can hardly be distinguished
from the ones obtained within the conventional convolution approach
of Ref. ~\cite{ciofi2}
(eq. \ref{pgn1}). According to the conclusions of Section 4.3 this
means
that relativistic effects (terms of order larger than ${p^2 \over
m_N^2}$)
are small, as also demonstrated in Ref.~\cite{coe}.
Our results fully confirm what found in \cite{ciofi2}, namely that
nuclear
structure effects in DIS of polarized electrons off polarized
$^3He$ are those due to the effective proton and neutron
polarizations
generated by the $S'$ and $D$ waves of $^3He$ wave functions, so that
the relation
 \begin{equation}
 g_1^{^3He}(x)\,\approx\,2p_p\,g_1^p(x)+ p_n\,g_1^n(x),
 \label{pg65}
 \end{equation}
$ p_p = -0.030$ and $p_n=0.88$ being
the effective nucleon polarization,
represents a reliable approximation of eq. (\ref{pgn1}) at
 $x\le 0.6$. The smallness of the difference between the free neutron
 structure function $g_1^n(x)$ and $g_1^{^3He}(x)$ is due to the
 smallness of the nuclear structure effects and is largely
independent of the
 form of the chosen parameterization for the nucleon SSF.
 For instance, using the
 parameters from the Ref.~\cite{shaef}, the results presented in Fig.
5 and
6 change by 20\%, i.e. by a quantity well below the experimental
errors,
 hence, eq.~(\ref{pg52}) may be considered as a good
 approximation for the extraction of the neutron SSF.

\section{Concluding remarks}
The necessity of plausible and precise data on the neutron
spin-dependent structure function is obvious. Since the information
about the internal neutron structure is predominantly obtained
from nuclear data, usually from the polarized deuteron and $^3He$,
an appropriate nuclear model for subtracting of the effects of
nuclear structure is requested.

In this paper a theoretical approach for the
analysis of polarized deep inelastic scattering
off light nuclei was proposed, which allows a self consistent
consideration of the role of the Fermi motion and the
meson degrees of freedom.
Since our model relays on the operator
product expansion method within OBE approximation, the $n-$th
moment of nuclear SSF have been
found as a product of two $n-$dependent functions. The contribution
of the impulse approximation and of the nuclear corrections to the
moments
have been separated in an explicit form. As a consequence, the
inverse Mellin transform
yields back
the nuclear SSF in a convolution form
with {\em two} distribution functions: one of them
describes the electromagnetic interaction of the lepton with an
on--shell nucleon, whereas the other one describes the strong
interaction of the hit nucleon with the other nucleons of the
nucleus.

A generalization of the model to the case of heavy
nuclei has been proposed and the comparison with the convolution
approach
based upon lepton scattering off bound off--shell nucleons
\cite{ciofi3} was performed
with the result that up to order $p^2/m_N^2$ the two approaches
coincide.
The numerical  calculations of  the spin-dependent structure
functions
for polarized deuteron and $^3He$ show that the nuclear corrections
are relatively small and essentially depend on the spin-orbital
structure of the corresponding nucleus, in agreement with the finding
of Refs.~~\cite{ciofi1}, based upon the conventional
convolution
approach.
For the deuteron the main
effect of the nuclear structure is due to  the destructive
role of the orbital motion of nucleons with $L=2$  and is of the
order of  magnitude $\sim\left( 1-3/2 P_D\right)$.
The comparison with the SMC experimental
results~\cite{smc} shows a reasonable agreement of our
calculations with the data.
In case of polarized $^3He$ the main nuclear structure
corrections come from the $S'-$ and $D$-waves admixtures
to the ground state wave function. Since they lead to a
partial depolarization of the neutron inside polarized $^3He$
and to an effective proton polarization,
$g_1^{^3He}$ slightly
differs from the free neutron spin-dependent structure function
$g_1^{n}$. The binding effects in both cases are found to be
small.
These results agree with those obtained within
the convolution approach\cite{ciofi1,ciofi2,kubj}.

To sum up, we can conclude that the convolution approaches so far
proposed
describe fairly well the peculiarities of polarized DIS off polarized
nuclei, so that nuclear corrections can be estimated in a reliable
way.
In closing, we would like to point out that our method based upon
the OPE within the effective meson nucleon theory, allows one to
microscopically understand the origin of the binding effect which is
present in the convolution approach. Moreover, since the expressions
for moments have been obtained explicitly, this allows one to
estimate the influence of nuclear effects on the Bjorken sum rule via
direct calculations of the first  moments.

\acknowledgments
We would like to thank Drs. A. Efremov, S. Gerasimov,
S. Mikhailov, E. Pace, G. Salm\`e
and O. Teryaev for fruitful discussions. Two of the
authors
(L.P.K. and A.Yu.U.) would like to thank INFN, Sezione di Perugia, for
warm hospitality and financial support.

\eject

\eject
\begin{figure}
\caption{The dressing diagrams for the helicity
amplitude corresponding to the forward Compton
scattering off the proton.
Black dots
denote the
scattering off a bare nucleon and the dashed line the pion
propagator. The first diagram is the contribution of the
impulse approximation
and renormalization terms, the second and diagrams
represent the recoil
effects and the last two diagrams are the self-energy-like
interaction
terms.}
\label{fig1}
\end{figure}

\begin{figure}
\caption{
The  helicity amplitude for the forward Compton
scattering off the deuteron. The notations are the same as in
Figure 1. The first three diagrams represent the impulse
approximation
with physical nucleons plus renormalization effects and the remaining
 two diagrams are the deuteron recoil and interaction terms,
respectively.}
\label{fig2}
\end{figure}

\begin{figure}
\caption{
The ratio of the deuteron and the isoscalar nucleon spin-dependent
structure
functions (SSF). Curve 1 : the impulse approximation;
curve 2: impulse approximation plus interaction term $\langle\{
S_z,V_\pi\}\rangle$.
Both curves have been computed using the parameterization of the nucleon
SSF
from Ref. \protect\cite{shaef};
Curve 3: the same as in curve 2, with the parameterization of
 the nucleon SSF from
Ref. \protect\cite{dor}; Curve 4: the same as in curve 2, plus the
contribution from the interaction
term $\langle\{ S_z,V_{OBE}\}\rangle$
representing the total contribution af all mesons
considered in the One Boson Exchange (OBE) interaction.
The arrow indicates the value
$(1-\frac{3}{2}P_D)$ (see text).}
\label{fig3}
\end{figure}

\begin{figure}
\caption{
(a) The weighted  deuteron spin-dependent structure function
$x g_1^D(x)$; (b) the first moment
${\sf M}_1(g_1^D)=\int_{x_{min}}^{m_D/m_N} g_1^D(x) dx$ versus the
lower limit of integration $x_{min}$.
The dotted (full) line corresponds to calculations with the
parameterization of the nucleon SSF
from Ref. \protect\cite{shaef}
(Ref. \protect\cite{dor}).
Experimental data from Ref. \protect\cite{smc}.}
\label{fig4}
\end{figure}

\begin{figure}
\caption{
The ratio of the $3^He$ and free neutron SSF calculated within the
impulse approximation (dotted) and within the impulse approximation plus the
interaction term $\langle\{ S_z,V_{OBE}\}\rangle$ (full). Curves 1 (2):
parameterization of the nucleon SSF
from Ref. \protect\cite{shaef}
(Ref. \protect\cite{dor}).}
\label{fig5}
\end{figure}

\begin{figure}
\caption{
The SSF for
$^3He$
calculated using the parameterization of the neutron SSF from
Ref. \protect\cite{shaef}, given by the dotted line.
Dashed line: the impulse approximation;
full line: the contribution
of the impulse approximation plus binding effects.
}
\label{fig6}
\end{figure}

\end{document}